\documentclass[11pt,a4,thmsb]{article}
\pdfoutput=1

\usepackage[letterpaper, margin=1in]{geometry}
\usepackage[toc,page,title,titletoc,header]{appendix}
\usepackage{fancyhdr}
\usepackage{caption}
\usepackage{subcaption}
\usepackage{graphicx}
\usepackage[T1]{fontenc}
\usepackage{units}
\usepackage{amsmath}
\usepackage{mathtools}
\usepackage{multirow}
\usepackage{array}
\usepackage{enumitem}
\usepackage{amssymb}
\usepackage{mathrsfs}
\usepackage{bm}
\usepackage{bbm}
\usepackage[margin=1cm]{caption}
\usepackage{pgfplots}
\usepackage{color, soul}
\usepackage{booktabs}
\usepackage{hyperref}

\usepackage[labelfont=bf]{caption}
\usepackage{threeparttable}
\usepackage{dcolumn}
\newcolumntype{d}[1]{D..{#1}} 

\begin{document}

\newcommand{\BetaD}{\mathtt{Beta}}
\newcommand{\Bern}{\mathtt{Bern}}
\newcommand{\Var}{\mathrm{Var}}  
\newcommand*\dif{\mathop{}\!\mathrm{d}}  


\title{\textbf{Battling Antibiotic Resistance: \\ Can Machine Learning Improve Prescribing?}\thanks{We benefited from very helpful feedback by Rolf Magnus Arpi, Lars Bjerrum, Gloria Cristina Cordoba Currea, Greg Crawford, Tomaso Duso, G\"unter Hitsch, Ulrich Kaiser, Jenny Dahl Knudsen, Sidsel Kyst, Chlo\'e Michel, Jeanine Mikl\'os-Thal, Maria Polyakova, Sherri Rose, Stephen Ryan, Karl Schmedders, Aaron Schwartz, Andr\'e Veiga, participants at the Annual Health Econometrics Workshop 2018 at Johns Hopkins University, the 2019 CESifo Area Conference on the Economics of Digitization, the Digital Economy Workshop 2019 at Cat\'{o}lica Lisbon, as well as in seminars at DIW Berlin, ESMT Berlin, University of Copenhagen, University of Zurich, and Vienna University of Economics and Business. Financial support from the European Research Council (ERC) under the European Union's Horizon 2020 research and innovation programme (grant agreement no. 802450) is gratefully acknowledged.}}

\author{\textbf{Michael Allan Ribers}\thanks{DIW Berlin and University of Copenhagen, Department of Economics - michael.ribers@econ.ku.dk}
\and \textbf{Hannes Ullrich}\thanks{DIW Berlin, University of Copenhagen (Department of Economics), University of Zurich (Department of Business Administration), Berlin Centre for Consumer Policies (BCCP), and CESifo - hullrich@diw.de.}}
\date{30 April 2019}

\maketitle

\vspace{-0.8cm}

\thispagestyle{empty} \setcounter{page}{0}

\begin{abstract}

\noindent Antibiotic resistance constitutes a major health threat. Predicting bacterial causes of infections is key to reducing antibiotic misuse, a leading driver of antibiotic resistance. We train a machine learning algorithm on administrative and microbiological laboratory data from Denmark to predict diagnostic test outcomes for urinary tract infections. Based on predictions, we develop policies to improve prescribing in primary care, highlighting the relevance of physician expertise and policy implementation when patient distributions vary over time. The proposed policies delay antibiotic prescriptions for some patients until test results are known and give them instantly to others. We find that machine learning can reduce antibiotic use by 7.42 percent without reducing the number of treated bacterial infections. As Denmark is one of the most conservative countries in terms of antibiotic use, this result is likely to be a lower bound of what can be achieved elsewhere.

\end{abstract}


\newpage

\section{Introduction}
Machine learning methods and the increasing availability of high-quality, large-scale data provide many new opportunities to design welfare improving policies with prediction problems at their core (Kleinberg et al.~2015, Agrawal et al.~2018, Athey 2018). The alarming increase in antibiotic resistance worldwide constitutes one such problem. In the US alone, antibiotic-resistant infections result in an estimated 23,000 deaths, \$20 billion in direct healthcare costs, and \$35 billion in lost productivity each year (CDC 2013).\footnote{Antibiotics are used to treat bacterial infections by killing or inhibiting growth of bacteria in the body. Their effectiveness is decreasing due to antibiotic resistance threatening to render simple infections, such as pneumonia or infections in wounds, a fatal risk. The World Health Organization considers antibiotic resistance one of the greatest threats to global health given its immense costs in terms of life and treatment expenses (WHO 2012, 2014).}
Human consumption of antibiotics is among the main drivers of antibiotic resistance. Therefore, limiting misuse of antibiotics is of crucial importance in curbing the development of antibiotic resistance. Correct prediction of bacterial vs.~other causes of infections is a key difficulty in treatment decisions. Machine learning combined with prediction-based policies has the potential to assist with this challenging task and reduce misuse of antibiotics.

When a patient first enters a physician office with symptoms of an infection, diagnostic information to inform the treatment decision is limited. The quality of treatment then depends largely on physician expertise. Such expertise includes the effective use of diagnostics, the interpretation of reported symptoms, and being able to handle potentially large amounts of background and medical data about each individual patient. In these tasks, physicians acquire information that is typically not encoded in data available to policy makers, making it difficult to evaluate physicians' prescribing practice. On the other hand, physicians may not have access to patients' full background data or lack the time or skills to analyze data for each individual patient. It is therefore an open question whether machine learning predictions based on administrative data can inform health policy to improve on decisions made by expert physicians.

In this paper, we propose prescription policies based on machine learning predictions of bacterial infection causes and evaluate improvements over expert general practitioner treatment decisions for patients suspected of urinary tract infection (UTI). UTI is one of the most common types of infections requiring antibiotic treatment and can be diagnosed with relative ease by microbiological analysis of urine samples.\footnote{Foxman (2002) reports almost half of all women contract a UTI once in their lifetime. In the United States, yearly UTI-related healthcare costs including workplace absences are estimated at \$3.5 billion (Flores-Mireles et al.~2015). According to Bjerrum and Lind\ae k (2015), each year 10 percent of women receive antibiotic treatment for UTI.} However, laboratory testing has the important and general limitation that results arrive with a delay of several days, corresponding to nearly a complete course of antibiotic treatment. Machine learning predictions promise a so far unavailable instantaneous bacterial risk assessment. We train a machine learning algorithm, a random forest, on high-dimensional, administrative data from Denmark in 2010-2012 to predict the risk of bacterial presence in laboratory test results from patient samples in primary care, the main source of antibiotic prescribing.\footnote{In the US, nearly half of all antibiotic prescriptions are made by primary care physicians (CDC 2015). In Europe, primary care accounts for 90 percent of prescriptions (Llor and Bjerrum 2014). In Denmark, considered here, general practitioners are responsible for roughly 75 percent of antibiotic prescriptions (Danish Ministry of Health 2017). While slightly imprecise we will use ``physician," "general practitioner," and "primary care physician" interchangeably.} The outcome variable, an indicator variable taking the value of one when bacteria are isolated in a patient sample, is based on the microbiological test result physicians receive several days after sending in a sample. The covariates in the prediction model include each individual patient's medical outpatient claims histories, past antibiotic prescriptions, past microbiological test results, a rich set of personal characteristics such as gender, age, detailed employment status and type, education, income, civil status and more, as well as the same information on each individuals' household members. We find that machine learning applied to these data is highly capable of predicting realizations of bacterial UTI in out of sample patient test results.

The relevant criterion on which our predictions need to be evaluated, however, is whether or not they can be used to improve human expert decision making. For this purpose, we model prescription decisions as a trade-off between the social cost of prescribing, i.e. promoting resistance, and the health benefits of antibiotic treatment. We build on Kleinberg et al. (2018), who use machine learning predicted risk of defendants committing a crime to show the potential improvements of judges' bail decisions. The model allows us to evaluate reassignment of antibiotic treatment based on the algorithm's prediction of risk, that is, delaying prescriptions for low risk patients until test results are available and instantly giving prescriptions to high risk patients. This reassignment is similar in spirit to Currie and MacLeod (2017) who evaluate the counterfactual of a reduction of C-sections for low-risk and an increase for high-risk pregnancies. While they focus on the effects of improving expertise, specifically decision making and surgical skills, we focus on the potential to use data-driven predictions to improve upon such expertise.

One important challenge in implementing prediction-based policies is that the patient sample to which a policy rule is applied in a real-world application varies over time and is unknown \emph{ex ante}. Sample variation does not only affect prediction quality. Even if the quality of the prediction function is unaffected by sample variation, the risk distribution varies with the sample of individuals arriving in clinics. If the policy rule depends on this distribution, as is the case in our application, then the policy rule risks missing the objective. To our knowledge, the current literature has performed counterfactual policy evaluation of machine learning predictions \emph{ex post} (Bayati et al. 2014, Kleinberg et al. 2018). An exception is work using randomized controlled experiments such as Dub\'e and Misra (2017) who show that a price targeting scheme successfully improves firm profits out of sample. To gain confidence that a policy objective can be attained using only \emph{ex ante} defined parameters, an additional step in the analysis is needed: continuous updating of the policy parameters prior to application of the policy rule. 

We propose a procedure relying on a three-way split of the data. The first partition is used to train the prediction algorithm, the second to fix the parameters of the policy function, and the third to evaluate counterfactual outcomes based on the policies defined in the first two steps. This procedure allows us to compare the improvements that can be achieved by defining the counterfactual based on \emph{ex ante} information with what would have been possible \emph{ex post}. Based on this comparison, we corroborate the cautious note by Rose (2018), who discusses the potential need to adjust prediction algorithms over time or other dimensions according to the specific context. The same is true for the specification of policies based on machine learning predictions, a subtle but important result of our paper.

We find that overall prescribing can be reduced by 7.42 percent without reducing the number of treated patients suffering from a bacterial infection. In comparison, the potential reduction in overall prescribing based on \emph{ex post} information is 9.57 percent. An alternative redistribution rule holds overall prescribing constant while aiming to increase antibiotic prescribing to patients with bacterial infections. We show that such a rule can improve the match between prescriptions and patients with bacterial infections by 6.38 percent. The potential increase in treated bacterial infections based on \emph{ex post} information is 7.87 percent. 

Importantly, for both objectives, we observe that 70 percent of patients to whom counterfactual policies give prescriptions at the initial consultation date receive prescriptions following the test result. Combined with an estimated 24 percent spontaneous recovery rate, this suggests that prescriptions based on machine learning predictions closely resemble physician choices under full information. We further provide evidence that physician choices are unlikely to be driven by heterogeneity in patients' health benefits from antibiotic treatment. Excluding a particularly vulnerable group, pregnant women, from the prediction-based policies, we find slightly lower policy-induced changes --- a 6.81 reduction in overall prescribing and a 5.56 increase in prescribing for bacterial infections --- with large overlaps of the confidence intervals of these effects. These observations suggest that we have indeed identified the potential to reduce wasteful prescribing by redistributing prescriptions from low-risk to high-risk patients. Finally, we highlight the importance of combining machine learning with physician expertise by evaluating a redistribution rule relying solely on machine learning predictions. Any rule that fails to include physician expertise cannot lead to improved prescription outcomes in the present context.

Comparing the distribution of predicted risk to physician prescription decisions, we find evidence consistent with some over- and underprescribing on average. Overall, primary care physicians prescribe very efficiently, reflecting the conservative culture towards antibiotic use in Denmark. Yet, conditional on predicted risk we find significant differences in prescription choices, suggesting heterogeneity in expertise. Physicians who correctly treat a large proportion of patients conditional on predicted risk likely have high expertise and make use of the data and diagnostic technologies available to them. We find that such expertise is positively correlated with clinic size and the likelihood to send test samples to the microbiological laboratory, and negatively with physician age. These observations hint towards stronger standardization and availability of diagnostic procedures in larger and younger offices with a higher aptitude towards diagnostic technologies and more exposure to UTI cases.

Using Danish data has two crucial advantages. First, Danish administrative data are unique in the world in scope and interconnectivity. They cover a vast array of information including patients' and patient household members' personal background information, detailed employment histories, as well as medical prescriptions and claims records, all of which are essential for conducting our analysis.\footnote{We find that similar results are achievable using a subset of predictors typically available to insurers or government agencies in developed countries. Yet, if that subset is too small, for example, containing only patient characteristics and prescription histories, the achievable reduction in prescriptions is significantly lower, around half of what we achieve with the full data.} In addition, unique person identifiers enable us to merge these data to individual laboratory test results acquired from two major hospitals in Denmark. Second, Denmark is a country with an excellent record of low antibiotic use (Goossens et al. 2005). In 2017, the Danish government initiated a national action plan in which one main goal is to reduce overall antibiotic prescribing by one third by 2020 compared to 2016 (Danish Ministry of Health 2017). Hence, the improvements we show for Denmark can be considered as a lower bound for the attainable improvement elsewhere.

Kleinberg et al. (2015) argue that prediction policy problems are important and commonplace. Existing work considers the prediction of shootings in Chicago to target a crime prevention program (Chandler et al. 2011), online reviews to predict hygiene inspections (Kang et al. 2013), prediction of household consumption responses for the targeting of a tax rebate program in Italy (Andini et al. 2018), and prediction of high-risk opioid prescriptions using administrative data from the US (Hastings et al. 2019). Kleinberg et al. (2018) use machine learning to evaluate the potential improvements of judges' bail decisions. In their setting, predicted crime rates for the jailed based on observables can be biased because only crimes committed by released defendants can be observed and judges might have selected these individuals based on unobservables. This problem of selective labels plagues many prediction problems. Kleinberg et al. (2018) propose a solution based on assumptions of random case assignment and varying leniency of judges, as well as homogenous risk prediction technologies across judges. In our setting, focussing on patients for whom physicians ordered laboratory test results at the initial consultation, the delayed arrival of diagnostic test results allows us to address the role of unobservables in physician decisions and to directly consider policy improvements. Focussing on this set of patients comes at the cost that the results cannot be extrapolated to the population of prescription decisions. Yet, in the context of antibiotic prescribing, our analysis holds empirical relevance covering nearly 100,000 consultations over a relatively short time period and small geographic area. Randomizing additional laboratory tests may be a feasible way to achieve further generalizable results.\footnote{Chalfin et al. (2016) predict worker productivity to improve police hiring practices and teacher tenure decisions, stressing the importance of decision makers' potentially complex payoff functions when drawing policy conclusions based on machine learning. For example, omitting important payoff dimensions, will lead to biased and potentially harmful policy outcomes. Medical decision making often has non-trivial payoff functions but antibiotic prescribing for UTI in primary care is comparatively straightforward. In our analysis, we investigate one potentially important payoff component, heterogeneity in patients' sickness disutility, by verifying that the policy outcomes are not significantly affected by dropping patients with high sickness disutility.} 

In medicine, machine learning for prediction has received much attention, driven by hopes that electronic health records, insurance claims, public registries, as well as genomics databases will help inform medical decision making. Obermeyer and Emanuel (2016) describe how machine learning will improve prognosis in clinical practice in the near future. For example, machine learning methods have been used and are expected to be used increasingly to predict organ failure or mortality, to automatize the interpretation of medical imaging such as mammograms, to automatize 24-hour monitoring in critical care, and to improve diagnostics. Chen and Asch (2017) highlight the challenges arising from the many complicating factors in the medical context. For example, predicting cancer treatment success in the distant future is a very difficult task based on historical health data. Google Flu Trends received much attention but had very little success when they attempted to predict the prevalence of influenza by combining large-scale online search data with a small sample of influenza cases. One core lesson from this low performance is that collecting and combining relevant types of data is crucial (Lazer et al. 2014). 

A large economic and public health literature considers demand-side mechanisms for policy interventions, including prescription surveillance and stewardship (Laxminarayan et al. 2013), general practitioner competition (Bennett et al. 2015), financial incentives for physicians (Yip et al. 2010, Currie et al. 2014, Das et al. 2016), education programs (Arnold and Straus 2005, Butler et al. 2012), peer effects (Kwon and Jun 2015), and social norm feedback (Hallsworth et al. 2016). To our knowledge, this paper is the first to directly address the core driver of wasteful prescribing: the lack of instant diagnostic information.


The remainder of the paper is organized as follows. Section 2 presents the institutional background and the data. Section 3 shows the results of the prediction algorithm. Section 4 presents the framework for the design and evaluation of prediction-based policy rules to improve antibiotic prescribing. Section 5 presents the achievable improvement in prescribing. Section 6 discusses potential threats to the validity of our results and Section 7 concludes.

\section{Danish Administrative Data and Laboratory Test Results}
We use Danish administrative data unique in the world in scope and interconnectivity. They cover a vast array of information including patient and patient household members' detailed socioeconomic data as well as antibiotic prescription histories, general practice insurance claims and hospitalizations, all of which are essential for conducting our analysis. In addition, the coherent use of unique personal identifiers enables us to merge these data to individual laboratory test results that we acquired from two major Danish hospitals. Before we explore the data in greater detail, we first give an overview of the institutional setting in Denmark, as Denmark has several regulations that impact general practitioner decision making that are therefore important for understanding the scope of our results outside the Danish context.


\subsection{The Danish healthcare system}
Denmark has a universal and tax financed single payer health care system with general practitioners as the primary gatekeepers. Every person living in Denmark is allocated to a general practitioner by a list-system within a fixed geographic radius around the home address. Patients can switch physicians from their initial assignment at a small cost but, effectively, most stick with their assigned general practitioner. Although general practitioners operate as privately owned businesses, all fees for services are collectively negotiated between the national union of general practitioners and the public health insurer and importantly, physicians do not generate earnings by prescribing drugs to patients who have to purchase their prescriptions from local pharmacies. Pharmacies earn a fixed fee per prescription processed regardless of the prescription drug price or other drug attributes, for example branded versus generic drugs. Prescription drugs are subsidized but patients co-pay a fraction of the list price depending on their cumulative yearly prescription drug expenditures. The Danish market for prescription drugs is highly regulated resulting in uniform pricing at pharmacies nationwide and antibiotic treatment is in general cheap, about 100 Danish Kroners (15 US Dollars) per complete treatment. General practitioners are responsible for prescribing approximately 75 percent of the human consumed systemic antibiotics in Denmark (Danish Ministry of Health 2017).

\subsection{Clinical microbiological laboratory test results}
For the period January 1st, 2010, to December 31st, 2012, we acquired test results from the clinical microbiological laboratories at Herlev hospital and Hvidovre hospital, two major hospitals in Denmark's capital region covering a catchment area of roughly 1.7 million people. The data contains patient and clinic identifiers and information on the test type, test acquisition date, sample arrival date at the laboratory, test result response date, isolated bacteria, and a list of antibiotic-specific resistances if bacteria were isolated. The laboratory test data are central because they reveal bacterial presence in a urine test sample, the outcome we aim to predict. The typical test procedure takes two to four days, during which general practitioners are uninformed about the test result. Since we know the precise timing of test acquisitions, prescription purchases, and the test response date, we can determine physicians' treatment decisions prior to being informed about test outcomes. Overall, the data contain 2,579,617 biological samples submitted for testing in the capital region by both general practitioner clinics and hospitals. Urine samples constitute 477,609 samples out of which 153,323 are submitted by general practitioners.\footnote{Urine samples from general practitioners are identified by department IDs UP1, UP2, UP3, and U-2, where the latter is crossed with general practitioner clinic codes as it contains both hospital and general practitioner samples.} We further restrict the number of test observations by excluding tests where the tested patient received a systemic antibiotic prescriptions or had another test conducted within 4 weeks prior to the respective test date. We make this restriction to focus on consultations that constitute a first contact with a physician within a patient's treatment spell. In these situations, physicians do not hold current test result information and must prescribe under uncertainty. By considering only initial consultations, we exclude potentially complicated treatment spells where patients are tested in later stages. We also avoid patients in longterm treatment, potentially due to severe antibiotic resistance problems. The resulting sample size we use for the analysis is 95,594 initial consultations, during which a sample was sent to a laboratory for testing. We define the test outcomes $y_{it}$ as an indicator for bacterial isolates where $i$ is a patient identifier and $t$ is the test acquisition time. Bacteria was isolated in approximately one out of three urine samples, both overall and among the general practitioner submitted samples. Importantly, the data on all 2,579,617 observed tests are used as input when training the machine learning algorithm as long as the observations are historical relative to the patient's test date.

\subsection{Danish national registries}
The administrative data provided by Statistics Denmark covers all citizens in Denmark between January 1, 2002, and December 31, 2012. For each individual, we observe a comprehensive set of socioeconomic and demographic variables, the complete prescription history of systemic antibiotics (\emph{L\ae gemiddeldatabasen}), hospitalizations (\emph{Landspatientregisteret}) and general practitioner insurance claims (\emph{Sygesikringsregisteret}). Information is linkable across registries via unique individual and physician identifiers.
The demographic data includes gender, age, education, occupation, income, marriage and family status, home municipality, immigration status and place of origin, and lastly includes household member identifiers. Hence, we can link family members and add demographic data for patient family members.
The data on systemic antibiotic prescriptions contain slightly more than 35 million purchased prescriptions. We observe the date of purchase, patient and prescribing physician identifiers, anatomical therapeutic chemical drug classification, drug name, price, indication of use, purchased package size and defined daily dose. The indication of use is imprecise in that many prescriptions are given with a UTI indication but prescriptions for UTI are also given with a more generic indication, e.g. against infection, or without indication at all.
The hospitalization data comprise all patient contacts with hospitals, including ambulatory visits. The data contain observations on hospitalizations of more than 2 million unique individuals per year over since 2002 and includes information on hospitalization admittance and discharge dates, procedures performed, type of hospitalization (ambulatory, emergency, etc), primary and secondary diagnoses and the number of total bed days.
Lastly, the insurance claims data cover all Danish general practitioner clinic services provided to the Danish population of patients. The claims data are comprised of approximately 100 million claims per year and includes physician and patient identifiers, consultation time, consultation, services provided and physician fees. A specifically important service contained in the claims data are in-house urine nitrate sticks tests which provide physicians with an indication, albeit imperfect, of whether UTI symptoms have a bacterial cause. We do not observe nitrate test results but observing claimed tests still gives us an indication of physicians private information relative to the machine learning algorithm. Further, the claims data allow us to identify pregnant women from mandatory pregnancy-associated examinations. This is important as pregnant women are by definition complicated UTI cases. We will discuss consequences of nitrate sticks test and the significance of pregnant women in detail following the presentation of our main results.
The combination of the laboratory data and the administrative registers yields a vector $x$ of 1,266 covariates that we feed into the prediction algorithm for each tested patient. All contained historic data relative to the test acquisition time are, in principle, observable to the physician at the time of the consultation.

\subsection{Physician prescribing relative to test results}
\begin{figure*}[h!]
    	\centering
    	\includegraphics[height=8cm]{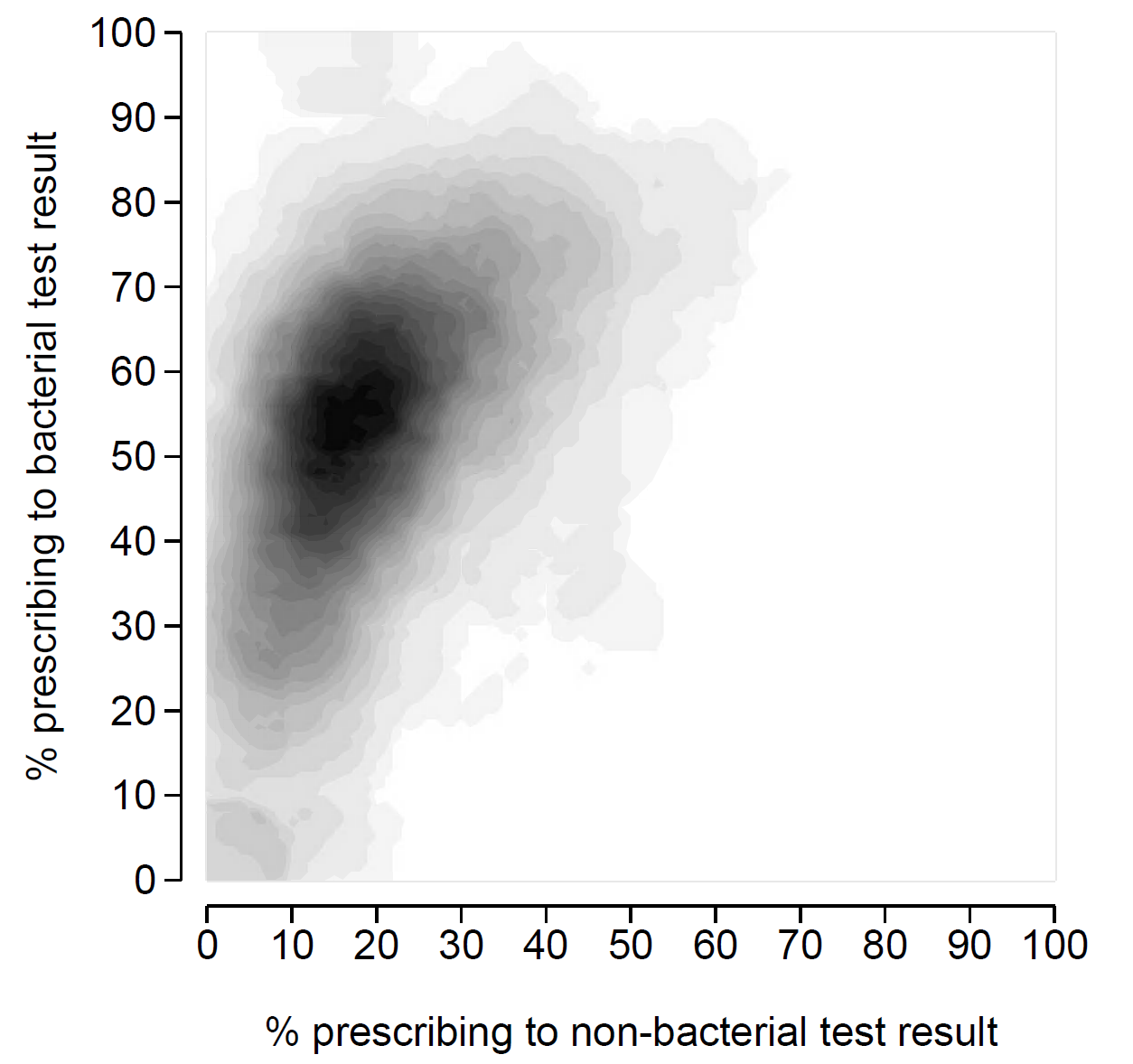}
	\caption{Heatmap of physician prescribing prior to knowing test results conditional on test outcomes. Due to anonymity restrictions, individual physicians cannot be showed.}
	\label{fig:prescr_bact_contour}
\end{figure*}

Before analyzing how machine prediction fares in identifying bacterial test outcomes, it is informative to inspect physicians' prescription choices relative to the outcomes. Prescriptions filed on the test acquisition day and up until observing the test outcomes reflect the physicians' preferences for antibiotic prescribing and their ability to identify bacterial infections but are of course relative to the selection of patients across clinics. Antibiotic treatment is known to be completely ineffective when bacteria are not isolated in tests and effective when bacteria are observed. Here we abstract from the fact that physicians need to choose one specific antibiotic drug out of several, not only whether to prescribe or not. Figure \ref{fig:prescr_bact_contour} shows a heatmap of the locations of physician clinics by prescription intensities relative to negative and positive bacterial test outcomes. To maintain anonymization required with the use of the administrative data, the heatmap only shows areas of three or more physicians. We can see that physician prescribing relative to bacterial outcomes varies widely. Physicians in the top left of the plot have high prescribing rates for bacterial infections and low prescribing rates for non-bacterial test outcomes. 
Overall, the plot suggests that general practitioners in Denmark do remarkably well in avoiding prescribing for non-bacterial outcomes while at the same time prescribing to a high share of bacterial-caused conditions UTIs. Yet, a non-negligible number of physicians located towards the top right, some of which cannot be plotted due to the anonymization restrictions. Physicians located to the top and right of a prescribing rates of (0.4,0.4) account for 8.4 percent of all physicians. This proportion is located in areas that indicate individual physicians' over- and underprescribing. Depending on the abilities of machine prediction in this context, improvements may be possible for all physicians, but the observed heterogeneity suggests that the possibility to detect over- and underprescribing will foster larger improvements for some physicians than for others.

\section{Machine learning}
Our empirical analysis consists of two steps. First, we train a random forest algorithm (Breiman 2001) that relates patient covariates to laboratory test outcomes in order to predict if patients suffer from \emph{bacterial} caused UTI and, consequently, if antibiotic treatment is beneficial or should be avoided. In a second step, we construct prediction-based prescription policy and evaluate counterfactual outcomes relative to observed physician prescription decisions. While the first step appears standard, our implementation differs from typical machine learning practice in that we do not randomly split our data in training and out of sample validation partitions. We intend to evaluate counterfactual analysis of prediction-based prescription policy and this requires that the prediction function and the policy rules are applied to future patients relative to the training data. Standard machine learning practice assumes that outcomes $y$ and covariates $x$ are independent draws from a joint distribution of $(Y,X)$ which remains the same for the training and out of sample partitions (Athey 2018). When the prediction function is applied to future patients and not random splits of the data, this assumption no longer holds by construction and we need to validate that our predictions and our policy rules remain valid over time. For this purpose, we create 24 monthly out-of-sample evaluation partitions from January 2011 to December 2012 and use all data prior to the respective test observations as training data. This section evaluates the random forest prediction quality while the following sections introduce our policy rules and evaluate out-of-sample policy results.
We do not dwell on the choice of machine learning method and certainly do not claim that the random forest algorithm outperforms other methods or that the out of sample predictions cannot be further optimized beyond what we accomplish here. The added value of our exercise comes from the ability to construct a sufficiently rich dataset that enables a \emph{standard} machine learning method to predict outcomes well enough such that we can implement a prediction-based policy that reduces wasteful prescribing. The core focus is on comparing machine predictions to physician antibiotic prescribing and, most importantly, on \emph{how} to construct and implement policy while remaining vigilant about the challenge with sample selection.

\subsection{Machine learning performance}
\label{sec:std_machine_performance}
\begin{figure*}
	\centering
	\begin{subfigure}[t]{0.46\textwidth}
    		\centering
    		\includegraphics[height=7.5cm]{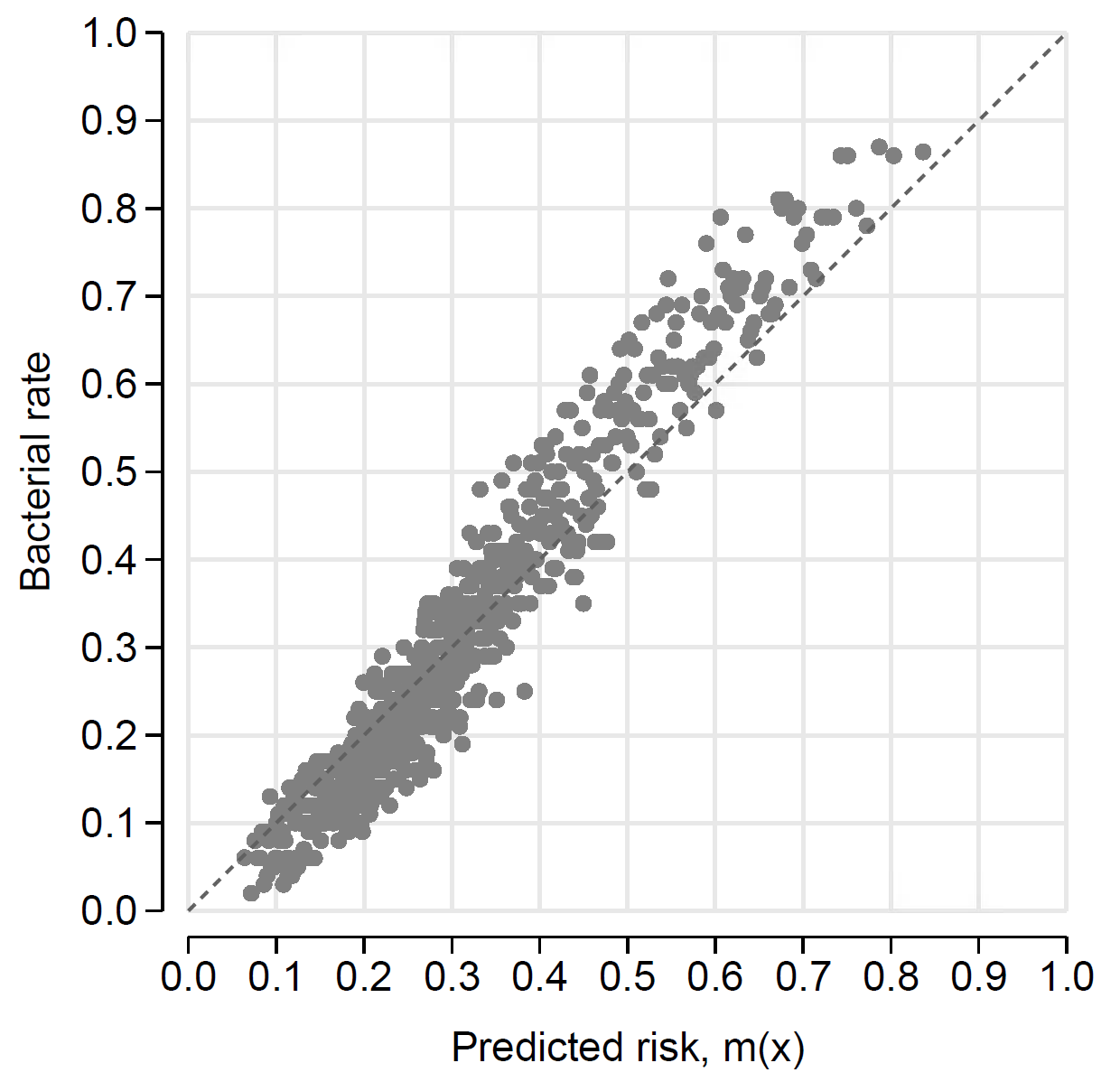}
		\caption{Laboratory test results relative to machine predictions of bacteria test outcomes. Each sphere represent averages over 100 tested patients sorted by the patients predicted risk of \emph{bacterial} UTI.}
	\end{subfigure}
	\hspace{0.5cm}
	\begin{subfigure}[t]{0.46\textwidth}
    		\centering
    		\includegraphics[height=7.5cm]{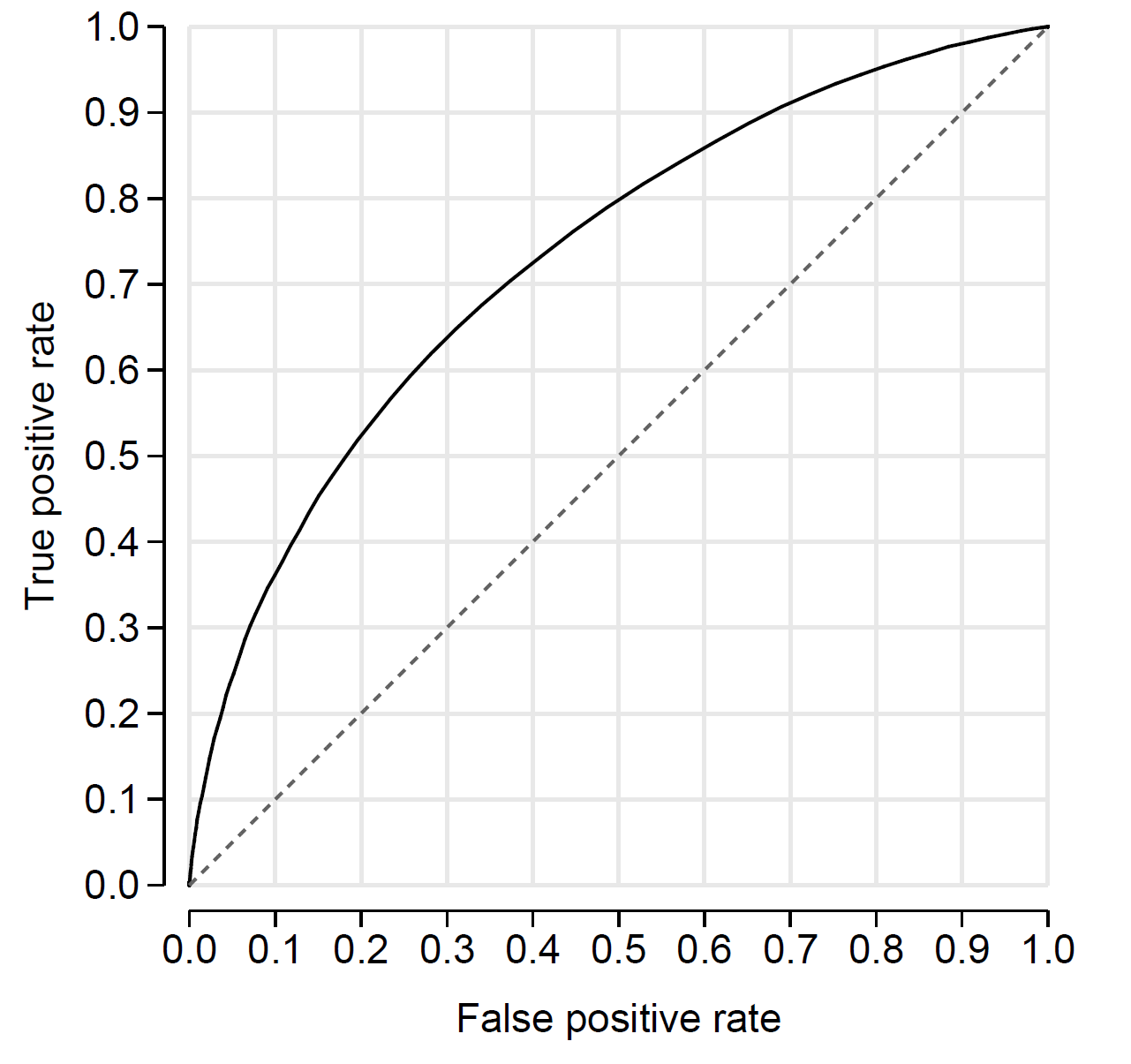}
		\caption{The receiver operating curve plotting the trade-off between the true positive rate and the false positive rate for varying classification thresholds. The area under the curve equals 0.731.}
	\end{subfigure}
	\caption{Prediction quality of the random forest algorithm out-of-sample.}
	\label{fig:bact_vs_predict}
\end{figure*}

A random forest is an ensemble of regression trees applied to bootstrapped versions of the training data. A tree represents a partition of the data created as a sequence of binary splits over individual variables where each split is determined by the homogeneity of the test outcomes in the created partitions. A simple model, in our case the mean, is universally fitted to all observations in each final partition, or leaf, of the tree. The random forest prediction, $m(x)$, is then the mean prediction over all trees.
We illustrate the prediction quality of our algorithm on the left panel of Figure \ref{fig:bact_vs_predict} which plots the average test results against the average algorithm predicted risk for all test observations in the joint set of monthly evaluation partitions. Every sphere represents a bin containing 100 patients where patients are assigned to bins sorted by their predicted risk. Outcomes are close to the 45 degree line throughout the risk distribution, showing that the algorithm on average correctly predicts bacterial risk. Among the riskiest 100 patients in the evaluation partitions, 86 percent are tested positive for bacteria following the initial consultation with the physician. Equivalently, the observed bacterial UTI rate for the 100 least riskiest patients is 6 percent. That the random forest predictions performs well throughout the risk range is a first necessity for improvements to physician decision making.

Binary predictions can be accomplished by comparing the random forest prediction to a classification threshold. Typical machine learning applications sets the classification threshold to pick majority vote, i.e. a classification threshold at $0.5$. However, other classification thresholds might be appropriate depending on the final application. As is standard in machine learning applications, we plot the receiver operating curve (ROC) on the out-of-sample evaluation observations as shown in Figure \ref{fig:bact_vs_predict}(b). The receiver operating curve plots of the true positive rate against the false positive rate as the classification threshold is varied from 0 to 1.
The closer the receiver operating curve is to the top-left corner, the better the prediction quality. A common metric by which to measure overall prediction accuracy is therefore the area under the ROC, the AUC.
Our \emph{bacterial} UTI prediction function has an AUC equal to 0.731. Kleinberg et al. (2018) report a comparable AUC of 0.707 in the crime risk context.

It is instructive to see which variables are the most important predictors. Figure \ref{fig:feature-importance} in Appendix \ref{app:RF} shows the feature importance for each variable, computed as the decrease in prediction error when variable values are permuted. Variables with feature importance equal to or below zero are considered to have no impact on prediction quality. For exposition, we collect variables in groups containing (i) patient characteristics and consultation dates; (ii) patient past prescriptions; (iii) patient past resistance test results; (iv) patient past hospitalizations; (v) patient past general practice insurance claims; (vi) household members' past prescriptions; (vii) household members' past resistance test results; and (viii) household members' past hospitalizations. The most notable result is that patients' past test results do not appear to be important features. This is likely explained by strong correlation between past test results and observed past antibiotic treatment, which cannot be captured by the feature importance metric.

\subsection{Physician prescribing conditional on predicted risk}
\label{sec:prescribing_cond_risk}

\begin{figure*}[h!]
	\centering
	\begin{subfigure}[t]{0.46\textwidth}
		\centering
		\includegraphics[height=7.5cm]{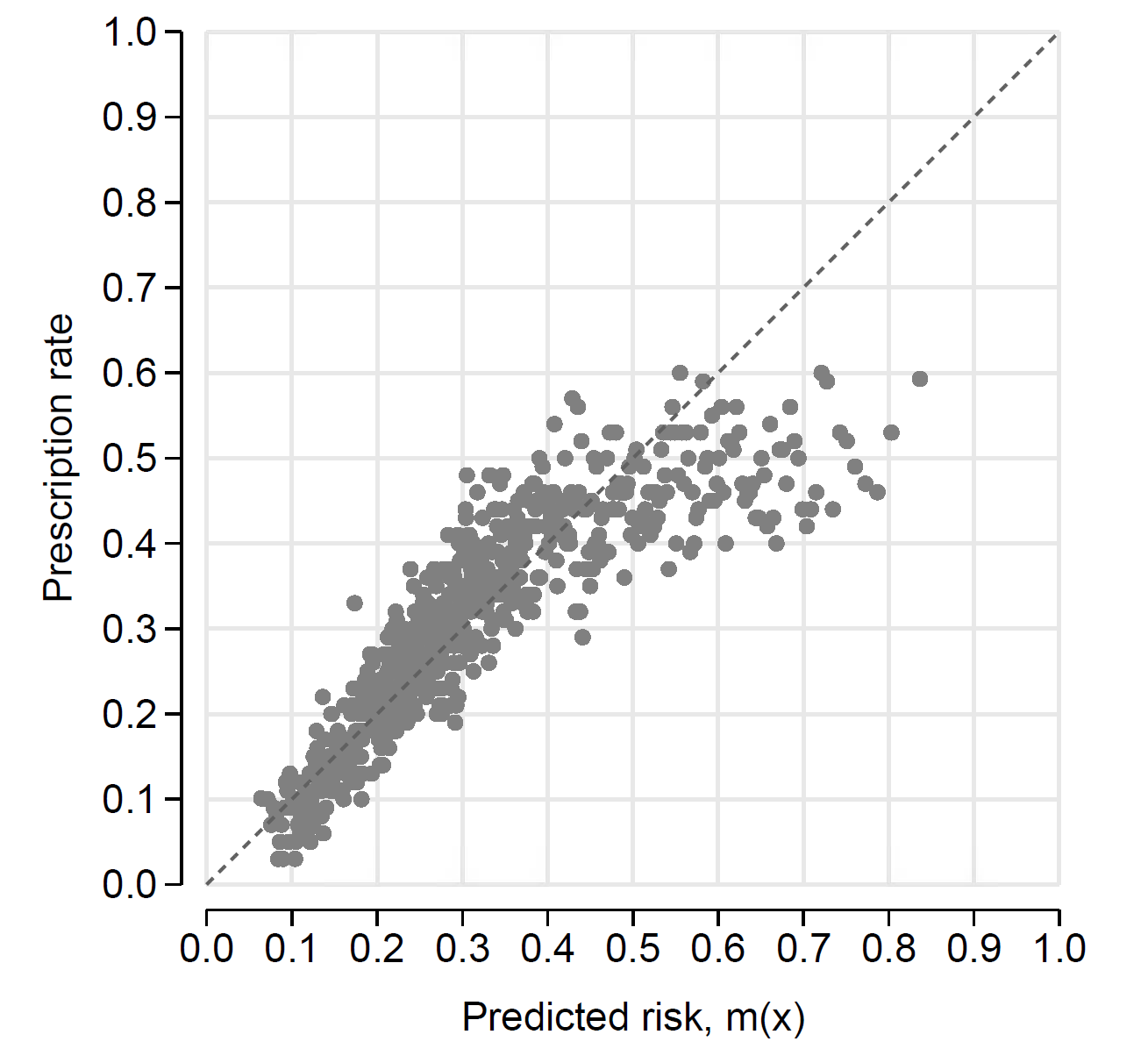}
		\caption{Machine predictions vs test outcomes conditional on antibiotic treatment prior to receiving test outcomes. Markers represent averages over bins of 100 patients sorted by predicted risk.}
	\end{subfigure}
	\hspace{0.5cm}
	\begin{subfigure}[t]{0.46\textwidth}
		\centering
		\includegraphics[height=7.5cm]{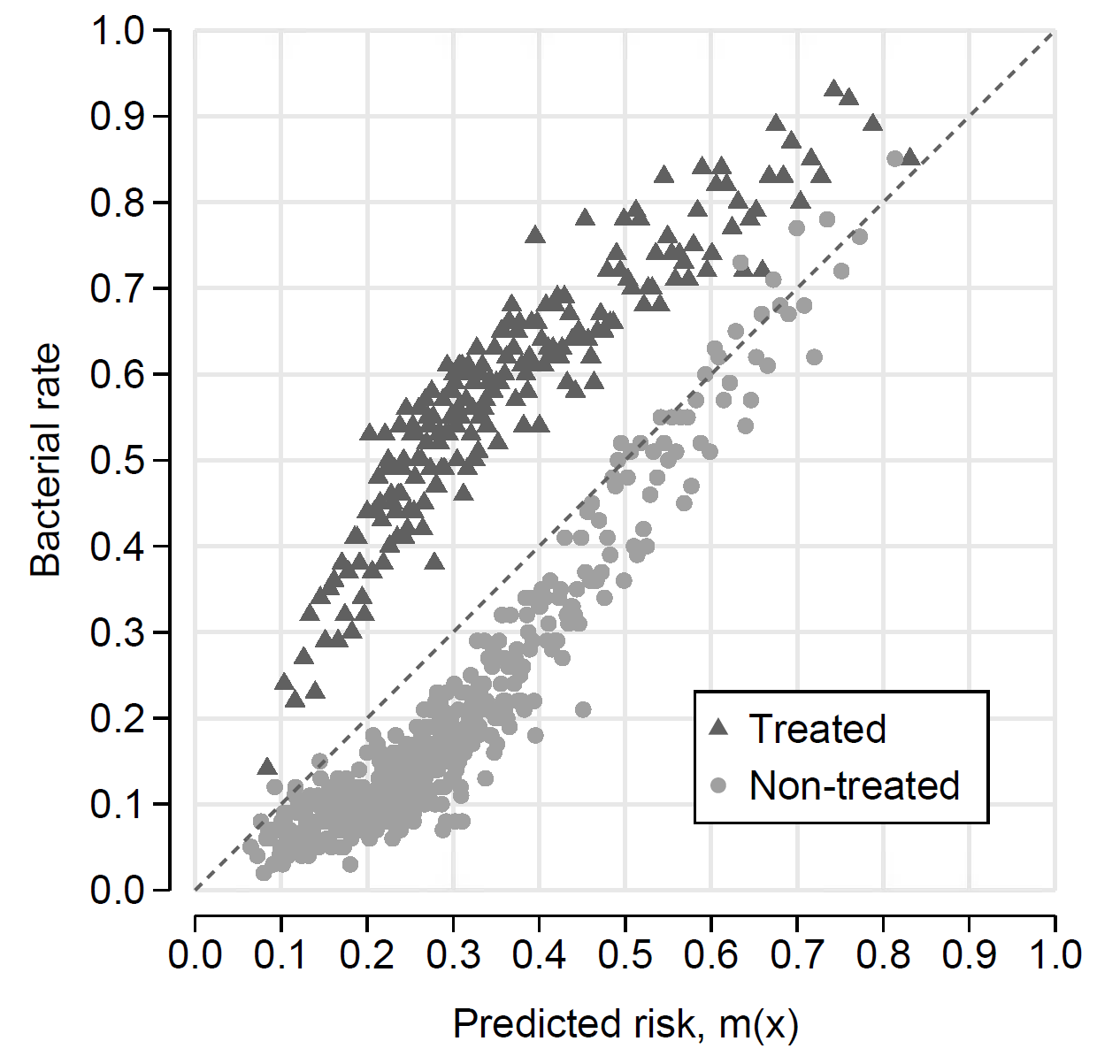}
		\caption{Mean deviation of bacterial rates and predicted risk, conditional on prescription choice. Markers represent averages over bins of 3 to maintain anonymity.}
	\end{subfigure}
	\caption{Bacterial test outcomes, prescribing, and unobservables.}
	\label{fig:prescribing_cond_choice}
\end{figure*}

The left panel of Figure \ref{fig:prescribing_cond_choice} plots the physicians' prescription rate prior to obtaining a test result against the algorithm's predicted risk. Again, spheres represent averages over bins containing 100 patients sorted by the algorithm's predicted risk. Physicians seem to evaluate low risk patients correctly on average, as the prescription rate and predicted risk appear well correlated in the low risk range. However, as predicted risk increases, the physicians' prescription rate flattens out, hence, the physicians and the random forest algorithm seem to disagree on the high risk patients who it appears that physicians have difficult time identifying. It is important to note that although physicians on average prescribe at the average bacterial rate, they are not always prescribing to the patients with bacterial UTIs, in fact far from it. To see this more clearly, we evaluate test outcomes versus algorithm predictions conditional on physician prescribing as shown in Figure~\ref{fig:prescribing_cond_choice}(b). It is important to note that patient predictions are not re-computed conditional on the prescription decision to prescribe but only re-sorted, i.e. physician instant prescriptions choices are not included in the covariates.
Several important insights can be observed. Conditional on the level of machine-predicted risk, physicians are on average able to prescribe to patients that more frequently show bacterial test realizations. This could be due to physician expertise and diagnostic unobservables, the latter, among other, covering in-house diagnostic tests such as nitrite dipsticks or microscopy.\footnote{Nitrite dipstick can detect bacteria that transform Nitrate to Nitrite. In the hold out data, the detectable genera are Escherichia, Enterobacter, Klebsiella, Citrobacter, and Proteus. The non-detectable genera are Staphylococcus, Pseudomonas, Enterococci, Acinetobacter, and Streptococcus. Inspecting prescription choices separately by dipstick-detectable and non-detectable bacterial species isolated in laboratory tests allows us to investigate whether physicians select on nitrite dipstick test results. While patients with dipstick-detectable bacteria have a higher prescription rate, 64 percent, relative to prescription rate for patients with non-dipstick-detectable bacteria, 55 percent, the difference is moderate. This suggests that dipstick test results leave significant uncertainty, which is consistent with evidence reported in the medical literature (Devill\'e et al. 2004).} The finding motivates that a prediction-based policy should include physician expertise for some range of predicted risk instead of relying solely on algorithm decisions. Figure~\ref{fig:prescribing_cond_choice}(b) also shows that physicians prescribe antibiotics to a substantial number of patients for whom predicted risk is very low, the lower left triangles. The 100 patients with the lowest predicted risk among the patients who received a prescription in the external partition on average had a positive bacterial test result in 13 percent of the cases. We define overprescribing as any prescription to a patient with a negative bacterial test result and observe that overprescribing on average decreases among the treated as machine predicted risk increases. Second, physicians do not prescribe to a substantial number of patients for whom predicted risk is very high, the top right spheres. The 100 riskiest patients not receiving prescriptions show a positive bacterial test result 84 percent of the times. We define underprescribing physician decisions to not prescribe to patients actually suffering from a bacterial infection and observe that underprescribing on average decreases as machine predicted risk decreases.

\subsection{Physician heterogeneity}
\begin{figure*}
\centering
		\includegraphics[height=7.5cm]{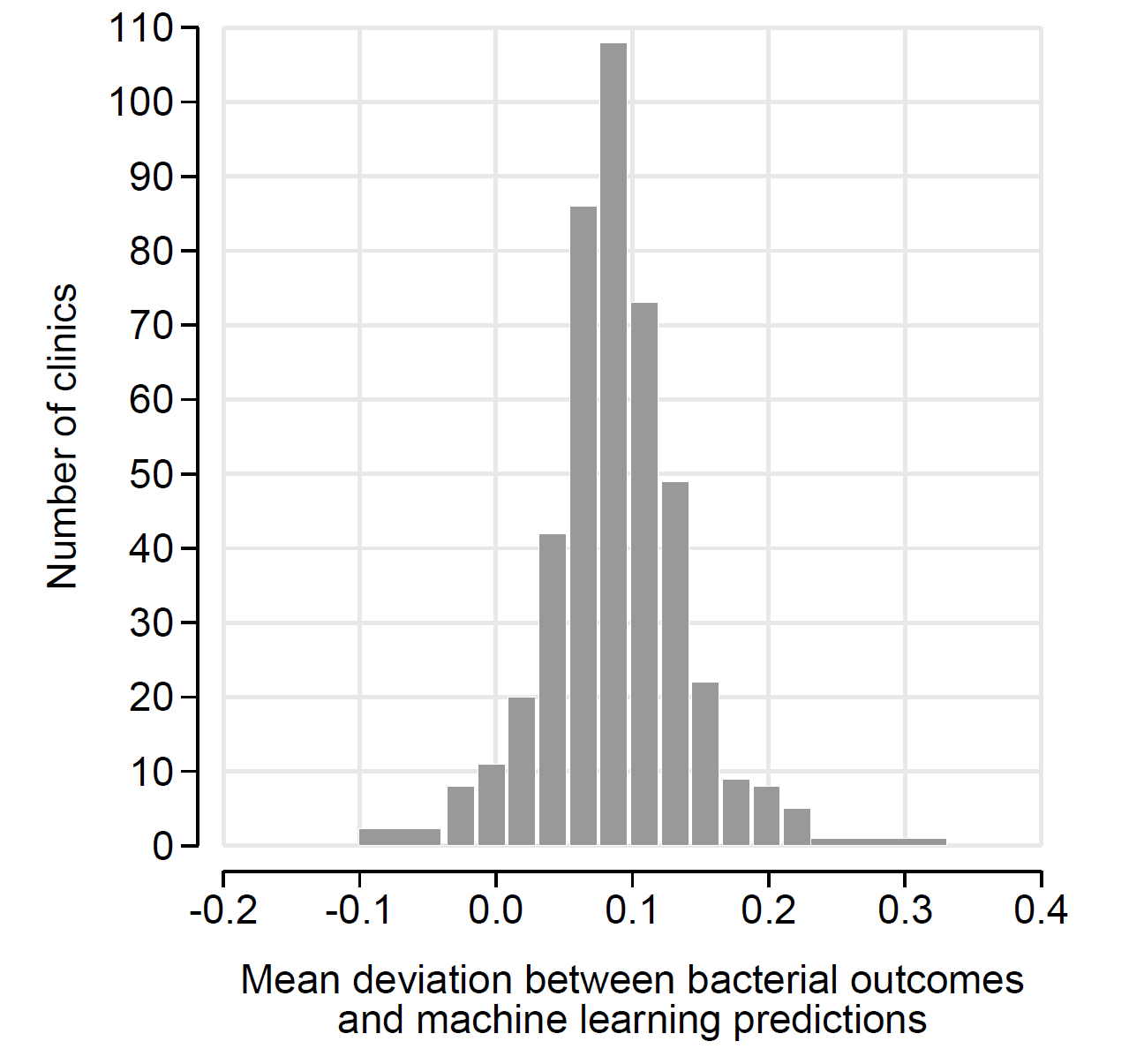}
		\caption{Distribution of mean deviation of bacterial rates and predicted risk conditional on prescription choice. The minimum bin size is set to three for anonymity.}
	\label{fig:meandeviation}
\end{figure*}

We find heterogeneity in physician expertise and use of (unobservable) diagnostics. Figure \ref{fig:meandeviation} shows the distribution of mean deviations between true bacterial outcomes and predicted risk conditional on treatment decisions at the physician level. This measure can be interpreted as the physician-specific distance between the bacterial rates conditional on predicted risk plotted in Figure \ref{fig:prescribing_cond_choice}(b). To preserve anonymity, each bin represents at least three physician clinics. Negative values can be read as physicians not separating bacterial from non-bacterial infections well conditional on risk. Physicians with positive values are able to separate bacterial from non-bacterial infections ``on top of" machine predicted risk. On average, the mean deviation is positive as can be seen in both Figures \ref{fig:prescribing_cond_choice}(b) and \ref{fig:meandeviation}. To learn about potential correlates with this measure, we link it to a set of clinic characteristics, which we observe for 265 out of the total of 451 clinics. Table \ref{tab:gp_meandev} in Appendix \ref{app:gp_meandev} shows the coefficients of a linear regression of the mean deviation on clinic characteristics. The number of patients per physician in a clinic is positively associated with physicians' ability to identify bacterial infections based on unobservables. One interpretation of this observation is that physicians with more frequent patient exposure are better at identifying bacterial infection causes. Another interpretation is that larger clinics may have better in-clinic equipment to gather same-day diagnostics during consultations. Both interpretations suggest heterogeneity in physician technologies for risk prediction. Physician age, on the other hand, is negatively associated, while the number of physicians in a clinic and the number of laboratory tests ordered per patient are positively associated with the ability to identify bacterial infections. While we can give no causal interpretation to the parameters estimated in this analysis, the correlations hint towards differences in the use of diagnostic technologies across clinics mainly based on clinic size, physician age, and the propensity to use laboratory diagnostics.

\section{Prediction-based prescription policy}
\label{section:machinepredpolicy}
The ability to predict bacterial test results does not necessarily hold value in itself. The relevant criterion by which our machine learning predictions need to be evaluated is whether or not they can be used to construct policy rules that improve the quality of physician prescribing.
To this end, we perform counterfactual analysis of prediction-based prescription rules implemented at patients' initial consultation of a potentially longer treatment spell. We restrict attention to observations where physicians submitted a urine sample for laboratory testing.
A common challenge when evaluating counterfactuals in machine learning applications is the selective labels problem.\footnote{In the bail decision context in Kleinberg et al.~(2018), the selective labels problem manifests itself in that crime outcomes are only observed for released defendants. If judges assess risk based on unobservables and make release decisions accordingly, crime risk for the jailed cannot be compared to crime risk of the released, even conditional on observables. Hence, any counterfactual release rule based on the released must be evaluated with care. One way around this issue is to evaluate a particular form of policy, namely release rules that jail similar defendants released by one judge but jailed by another. Two necessary assumptions for this approach are random assignment of defendants to judges and varying leniency in release decisions across judges.} By focusing on the sample of consultations that include microbiological testing, test outcomes are observed for all patients regardless of the physician's prescription decision. This means we do not face the selective labels problem and can evaluate antibiotic prescription rules directly, allowing us to construct and evaluate redistribution rules based on machine prediction. The disadvantage to our approach is that we cannot claim the generalizability of our results to prescription occasions that did not include patient microbiological testing. However, the tested cases are significant in number, accounting for approximately 15 percent of all initial UTI consultations. Further, when a physician decides to test, the value of diagnostic information is presumably high so that the prediction-based policies proposed here improve upon situations in which physicians are making decisions under significant uncertainty. In practice, policy makers push for more intensive use of laboratory testing for antibiotic prescription choices so that reducing the selection problem is feasible in future research and policy implementations.

\subsection{Policy rules}
We limit the set of considered policy rules in two ways. First, we only consider three actions: delay prescribing until test results are available, instantly prescribe prior to receiving test results, and no interference with physician prescription decisions. Second, actions are implemented conditional on algorithm predicted risk only. This means that we do not allow policy rules to directly discriminate on physician or patient characteristics; only indirectly through realizations of algorithm predicted risk. We do this to avoid the situation where a patient would be given a prescription by the algorithm when consulting one physician but not by consulting another.\\
\noindent
The observed prescription and outcome data inform a specific set of policies we will consider.
Specifically, in section \ref{sec:prescribing_cond_risk} we see that overprescribing occurs most frequently at low predicted risk and decreases on average as risk increases. Similarly, we see that underprescribing most frequently occurs at high predicted risk and decreases as predicted risk decreases.
From these observations, we infer that a prediction-based policy should delay prescriptions for patients with low predicted risk until test results are available, assign prescriptions prior to observing test results for patients with high predicted risk, and not interfere with physician prescription decisions in a middle range where the algorithm predictions are the least informative.
Hence, we will consider machine learning policies of the following form:
\begin{equation}
	\rho \left( m(x) ; \rho^{J} , k_{L}, k_{H} \right) \;\; = \;\;
		\begin{cases}
			\;\; 0		& \hspace{0.4cm} \text{if } m(x) < k_{L}, \\
			\;\; \rho^{J}	& \hspace{0.4cm} \text{if } k_{L} \leq m(x) \leq k_{H}, \\
			\;\; 1		& \hspace{0.4cm} \text{if } k_{H} < m(x),
		\end{cases}
\label{eqn:general_stewardship_rule}
\end{equation}
where $x$ is a vector of patient's explanatory variables at the time of laboratory test sample acquisition, $m(\cdot)$ is the random forest prediction function, $\rho^{J}$ is the physician's prescription choice and $(k_{L}, k_{H})$ are policy parameters to be determined. Such policy rules delay prescribing for a patient if the patient's bacterial risk is predicted below $k_{L}$ and assigns an instant prescription to a patient for whom the bacterial risk is predicted above $k_{H}$. Patients with bacterial risk between $k_{L}$ and $k_{H}$ are unaffected by the policy.

\subsection{Policy objective} \label{sec:rules}
We assume that a policy maker considers the prescription decision, $p$, at the first consultation as a trade-off between the patient's sickness cost, $a$, accrued from deferring prescribing until a test result is available, and the social cost of prescribing, $b$, that is, promoting antibiotic resistance.
While the social cost are incurred for every antibiotic prescription, prescribing an antibiotic does not necessarily alleviate the sickness cost. Antibiotics only have a curative effect if a patient suffers from a \emph{bacterial} infection. Patients with UTI symptoms related to non-bacterial causes are unaffected by antibiotic treatment and suffer their sickness cost regardless.\footnote{Allowing for differing sickness costs for a variety of non-bacterial ailments does not affect the model outcome. Our aim is to \emph{compare} antibiotic prescription policies and since antibiotic use only impacts health outcomes for patients suffering from bacterial UTI, patients with other ailments are unaffected by treatment and their sickness costs drop out in the final analysis.}
Hence, we model the policy maker's payoff at a patient's initial consultation by
\begin{equation}
	\pi(p; y)  = -ay(1-yp)-bp ,
\label{eqn_policy_payoff_function}
\end{equation}
where 
$y$ is the realization of the test outcome, i.e. an indicator for whether the patient's UTI is caused by bacteria. We assume that $0<b<a$ such that prescribing is always optimal when an infection is known to be bacterial with certainty.\footnote{Our application does not strictly require $b<a$, only that $0<a$ and $0<b$. However, if $a<b$, it is never optimal to prescribe prior to observing test results, which makes the whole exercise redundant and is not reflected in our data or the Danish antibiotic prescription guidelines (Danish Health and Medicines Authority 2013).} The test outcome is not observed at the time of the initial consultation. However, a policy maker can evaluate prescription quality in hindsight by computing differences in outcomes between the prediction-based policy in equation (\ref{eqn:general_stewardship_rule}) and physicians prescription choices:

\begin{align}
\begin{split}
	\Pi(k_{L},k_{H})
       	&= \mathop{E}_{x,y,\rho^{J}} \! \left[ \pi \! \left( \rho \!\left( m(x); \rho^{J} \! , k_{L} , k_{H} \right) \!, y\right)-\pi(\rho^{J} \!, y) \right] \\
	&= a  \underbracket{ \mathop{E}_{x,y,\rho^{J}} \! \left[ \, y \! \left( \rho \!\left( m(x); \rho^{J} \! , k_{L} , k_{H} \right) - \rho^{J} \right) \right] }_{ \Delta \text{bUTI}} \; 
	- \; b  \underbracket{ \mathop{E}_{x,\rho^{J}}  \left[ \rho \!\left( m(x); \rho^{J} \! , k_{L} , k_{H} \right) - \rho^{J} \right]}_{\Delta \rho} ,
\end{split}
\label{eqn:payoff_increase}
\end{align}
where the expectation is over realizations of $x$, $y$, and $\rho^{J}$ over the period of observation. The impact of the prediction-based policy can be separated into two terms: the benefit from the additional correctly treated bacterial UTI patients, $\Delta \text{bUTI}$, and the change to the number antibiotics used, $\Delta \rho$. Clearly, if the prediction policy increases the number treated with bacterial infections, $\Delta \text{bUTI}>0$, while reducing the number of antibiotic prescriptions, $\Delta \rho<0$, a policy maker is better off regardless of the values of $a$ and $b$. However, depending on the policy maker's preferences, it might be optimal to implement a prediction-based policy that targets a reduction in antibiotic prescribing at the expense of delaying treatment for an increasing number of patients; or equivalently, a policy maker might prefer to increase the number of correctly treated bacterial UTIs even if it means increasing overall antibiotic use. The rate at which the policy maker intends to do so depends on the relative size of $a$ and $b$ and the rate at which algorithm predictions can identify bacterial UTIs.
We do not observe policy maker preferences and therefore also do not know the trade-off a policy maker would prefer. Therefore, we employ a strategy inspired by Kleinberg et al.~(2018). We aim to find the largest possible reduction in antibiotic prescribing without changing the number of treated bacterial UTIs by setting policy thresholds according to:
\begin{equation}
	\min_{ k_{L} , \, k_{H} } \Delta \rho \quad s.t. \quad \Delta \text{bUTI} = 0 .
\label{eqn:AB_policy_rule}
\end{equation}
We refer to this policy objective as the \emph{antibiotic reduction rule}.
A second objective can be to increase the number of treated bacterial UTIs without changing the number of antibiotics used, i.e. to set policy thresholds such that:
\begin{equation}
	\max_{ k_{L} , \, k_{H} } \Delta \text{bUTI} \quad s.t. \quad \Delta \rho = 0
\label{eqn:bUTI_policy_rule}
\end{equation}
We refer to this policy objective as the \emph{increase treated bacterial UTI rule}. In Section \ref{sec: ex-ante vs ex-post}, we determine these optimal thresholds from known distributions and discuss how to estimate them \emph{ex ante} when the distribution of patients and physician decisions are unknown.

\section{Counterfactual policy outcomes}
\label{sec: ex-ante vs ex-post}
We begin by evaluating the \emph{ex post} potential of the policy rules in equations (\ref{eqn:AB_policy_rule}) and (\ref{eqn:bUTI_policy_rule}). Here, out-of-sample policy outcomes are evaluated for policy parameters $(k_{L}, k_{H})$ optimally set \emph{ex post}. Hence, we can assure that the policy constraints are met as the realizations of patient covariates, bacterial test outcomes, and physician prescription decisions are observed.
However, a real implementation of prediction-based policies requires the policy parameters to be determined \emph{ex ante}, i.e. $(k_{L}, k_{H})$ must be estimated prior to an out-of-sample evaluation period.
Sample variation and unknown time or seasonal trends can pose a serious challenge for implementations in practice. We propose an \emph{ex ante} approach that comes close to realizing the \emph{ex post} achievable outcomes while satisfying the constraints in equations (\ref{eqn:AB_policy_rule}) and (\ref{eqn:bUTI_policy_rule}).
To show the robustness of this approach in the context of antibiotic prescribing, we repeat the counterfactual implementations independently for each month throughout the sample period. 

For all implementations, \emph{ex ante} and \emph{ex post}, we present policy outcomes relative to observed levels. Hence, for the antibiotic reduction rule, we compute the percentage change in antibiotic use out of all observed initial antibiotics prescribed:
\begin{align}
	\% \Delta \rho &= \frac{ \mathop{E}_{x,y,\rho^{J}} \! \left[ \rho \!\left( m(x); \rho^{J} \! , k_{L} , k_{H} \right) - \rho^{J}   \right]}{ \mathop{E}_{\rho^{J}} \! \left[ \rho^{J} \right]} ,
\label{eqn:prescribing_percent}
\intertext{and equivalently, for increase treated rule, we compute the additional treated bacterial UTI patients out of all initially correctly treated:}
	\% \Delta \text{bUTI} &= \frac{ \mathop{E}_{x,y,\rho^{J}} \! \left[ \, y \! \left( \rho \!\left( m(x); \rho^{J} \! , k_{L} , k_{H} \right) - \rho^{J} \right) \right] }{\mathop{E}_{y,\rho^{J}} \! \left[ \, y \rho^{J} \right]} .
\label{eqn:bUTI_percent}
\end{align}
%

\subsection{\emph{Ex post} policy potential}
\label{sec: ex-post policy}

The evaluation of policies requires the specification of out-of-sample evaluation periods. The length of these evaluation periods can affect policy evaluation in important ways. If evaluation periods are long, uniform policy parameters might fail to achieve the full policy potential due to variation in the tested population over time. For example, in our context we observe an increasing number of patients over time. If the growing number of patients changes the bacterial risk distribution, then a single $(k_{L},k_{H})$ will not be optimal in the sense of the policy objective. If the evaluation periods are short, sample sizes can become small, resulting in large variation of policy outcomes over time and extreme changes to policy rules over time.
We determine \emph{ex post} policy parameters and evaluate policy outcomes at the monthly level as follows. For each evaluation month from January 2011 to December 2012, we use all past observations relative to the evaluation period as training data for the random forest algorithm. Based on the algorithm's predictions, we evaluate the out-of-sample policy potential \emph{ex post} for each evaluation month.\footnote{The training data therefore grows over time. For the evaluation period January 2011, the training data comprise all of 2010. For the evaluation period January 2012, the training data comprise 2010 and 2011. We have replicated our analysis training the random forest on data only including observations one year prior to the evaluation period, i.e. having all training partitions with equal length. This does not affect the conclusion of our results.}
\begin{figure*}
	\centering
	\begin{subfigure}[t]{1\textwidth}
		\centering
		\includegraphics[height=7.5cm]{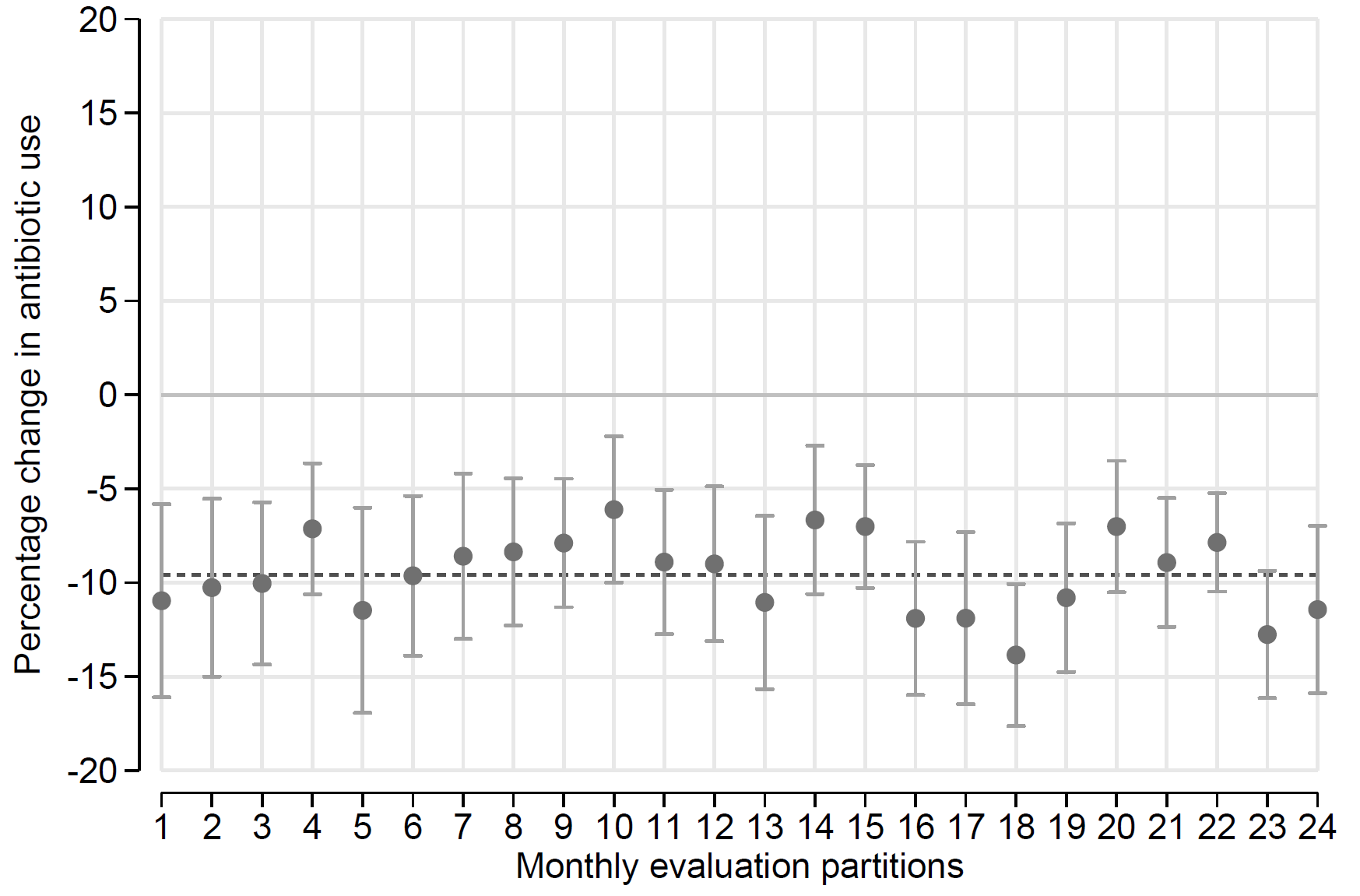}
		\caption{Antibiotic reduction}
	\end{subfigure}\\
	\vspace{1cm}
	\begin{subfigure}[t]{1\textwidth}
		\centering
		\includegraphics[height=7.5cm]{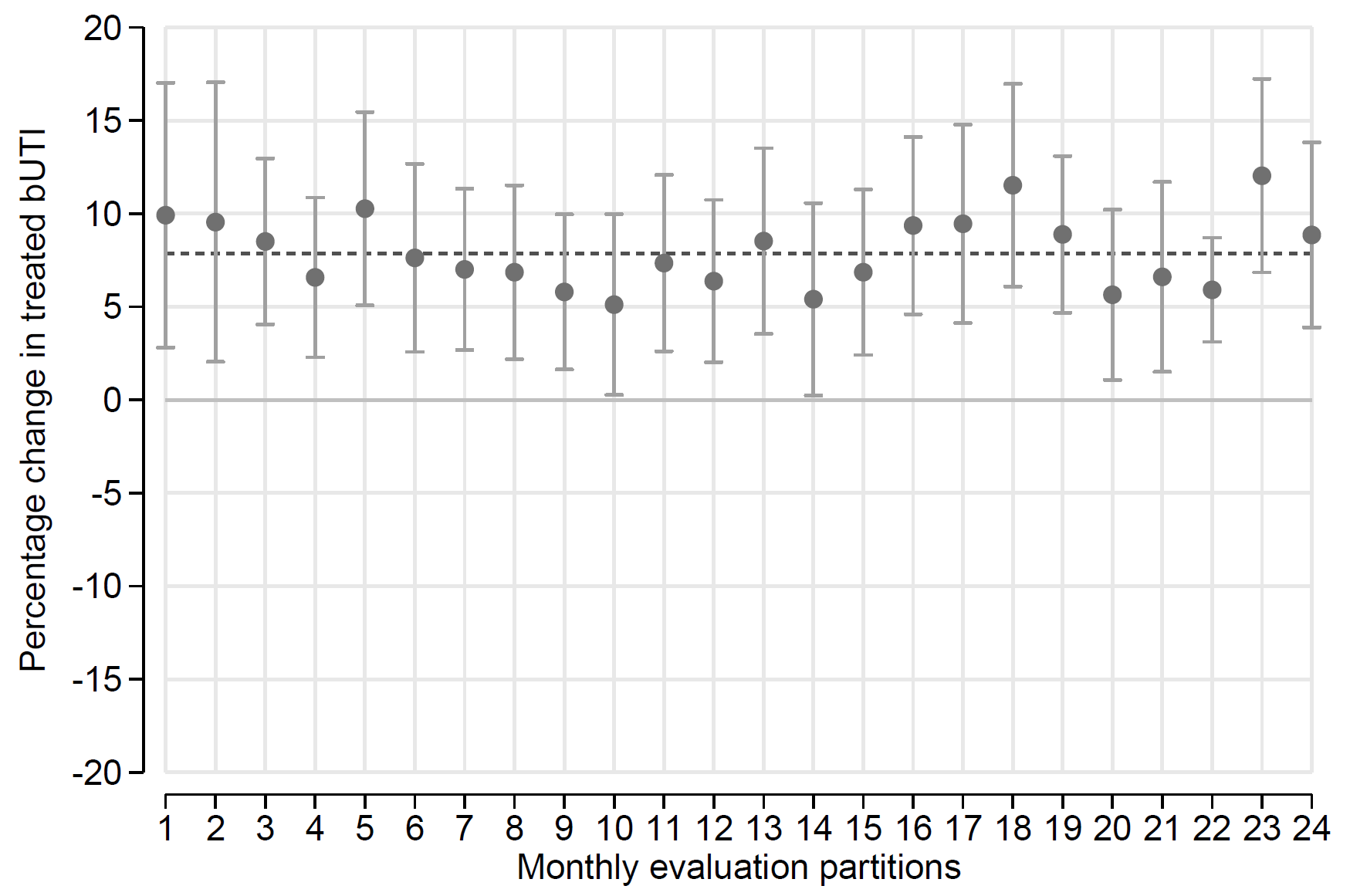}
		\caption{Increase in treated bacterial UTIs}
	\end{subfigure}
	\caption{\emph{Ex post} potential of each policy rule by monthly evaluation periods from January 2011 to December 2012. 95\% confidence intervals are computed by bootstrapping the evaluation periods 100 times. The dashed line is the outcome for all evaluation periods combined.}
	\label{fig:ex-post-policy}
\end{figure*}
Figure \ref{fig:ex-post-policy} shows the ex post policy potential for each evaluation period. Confidence intervals are generated by bootstrapping each evaluation period 100 times holding the \emph{ex post} policy parameters fixed.
The antibiotic reduction rule has the potential to reduce antibiotic use by about 10 percent in all months without treating fewer patients suffering from bacterial UTI. The average across the entire sample is a reduction of 9.57 percent. Each month shows variation in outcomes due to sample variation that affect policy outcomes even when policy parameters are chosen \emph{ex post}.
Equivalently, the increase treated rule achieves a 7.87 percent increase in treated patients suffering from bacterial UTI without increasing antibiotic use. The policy potential is significantly different from zero for all evaluation periods except for period 10 (October 2011) where the 95\% confidence interval just crosses zero.
Table \ref{tab:ex-post-policy} summarizes the out-of-sample \emph{ex post} policy outcomes aggregated over all evaluation periods.
Although the \emph{ex post} constraints are zero by definition on the individual evaluation periods, the constraint is violated when the \emph{ex post} policy parameters are applied to bootstrapped evaluation samples, hence, we provide bootstrapped confidence intervals for the constraint as well as the policy aim. Figure \ref{fig:ex-post-policy_constraint} in Appendix \ref{app:ex-post-violation} shows the monthly confidence intervals on the constraints.
\begin{table}[h!]
\centering
\small
\begin{threeparttable}
\caption{\emph{Ex post} policy outcomes}
\label{tab:ex-post-policy}
\begin{tabular}{ @{} l @{\extracolsep{10mm}} c @{\extracolsep{6mm}} c @{} } \\[-1cm]
\toprule
				& \multicolumn{2}{c}{Percentage change in} \\
				\cline{2-3}
Policy rule 		& Antibiotic use 	& Treated bacterial UTIs \\
\midrule
Antibiotic reduction		& $-9.57 \ \left[-10.38, -8.76\right]$ 	& $0  \ \left[-1.03, 1.03\right]$ \\
Increase treated bacterial UTI 	& $0 \ \left[-0.81, 0.81\right]$ 	& $7.87 \ \left[6.90, 8.85 \right]$ \\
\bottomrule
\end{tabular}
\vspace{-0.2cm}
\footnotesize 95\% confidence intervals are computed based on 100 bootstrap samples.
\end{threeparttable}
\end{table}

\subsection{\emph{Ex ante} policy outcomes}
\label{sec:ex-ante-policy}
The implementation of prediction-based policies in practice requires determining policy parameters prior to evaluation periods.
The challenge is to estimate policy parameters \emph{ex ante} that come close to realizing the \emph{ex post} policy outcome reported in section \ref{sec: ex-post policy} while ensuring that the constraints hold, i.e. that either antibiotic use or the number of treated bacterial UTI stays constant.
To compare the results to the benchmark \emph{ex post} outcomes above, we set the policy parameters with the aim to satisfy the constraints at the monthly level.
By evaluating the \emph{ex ante} policy outcome every month, we replicate our procedure 23 times (February 2011 through December 2012) which permits validation of the success rate of our approach. Depending on the implementation setting, constraints can be very sensitive to changes in the policy parameters so that small changes in distributions, for example due to time trends, can lead to large constraint violations. In our context, targeting the full potential of a policy leads to constraint violations in eight of 23 months and nine of 23 months for the antibiotic reduction rule and the treated bUTI rule, respectively. A convenient approach to shield from such constraint violations is to frequently update policy parameters and to implement conservative policies for which constraints are likely to hold. For this, we propose the following procedure:
\vspace{0.2cm}
\begin{center}
\begin{minipage}{14cm}
	\begin{itemize}
		\item[Step 1:] Train a prediction model at time $t$ using all past observations and predict test outcomes for an evaluation period of length $\tau$ starting at $t$.
		\item[Step 2:] Based on the predictions in the evaluation period, compute conservative $\hat{k}_{L}$ and $\hat{k}_{H}$ policy parameters that target a fraction $\alpha$ of the \emph{ex post} policy potential.
		\item[Step 3:] Set $\hat{k}_{L}$ and $\hat{k}_{H}$ as \emph{ex ante} policy parameters for a period of length $\lambda$ immediately following the \emph{ex post} evaluation period defined in Step 1.
		\item[Step 4:] Repeat steps 1 to 3 moving through time with step size $\lambda$ until the sample period is covered and evaluate policy outcomes based on the \emph{ex ante} $\hat{k}_{L}$ and $\hat{k}_{H}$ policy parameters.
	\end{itemize}
\end{minipage}
\end{center}
\vspace{0.3cm}

\begin{figure*}
	\centering
	\begin{subfigure}[t]{1\textwidth}
		\centering
		\includegraphics[height=7.5cm]{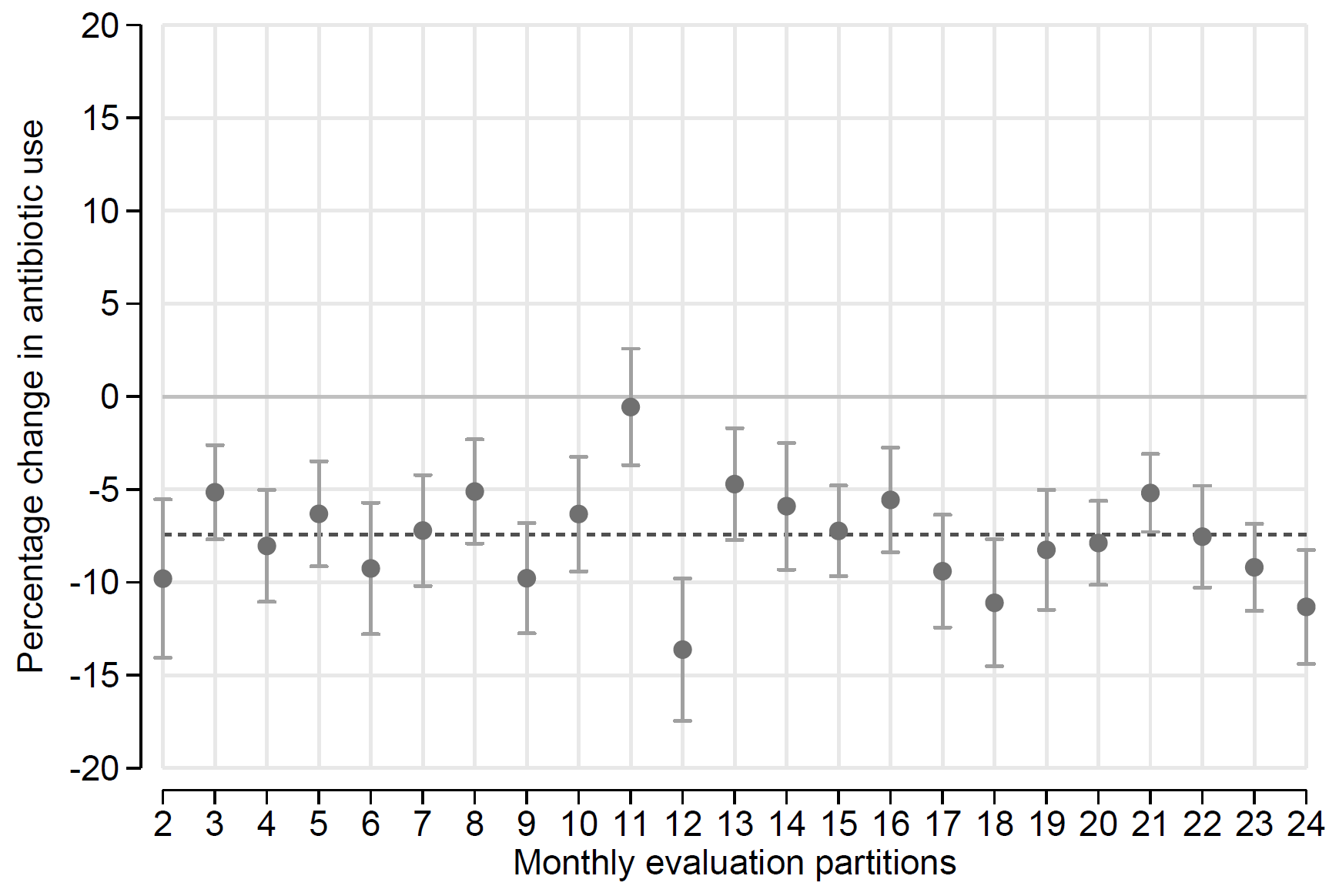}
		\caption{Antibiotic reduction}
	\end{subfigure}\\
	\vspace{1cm}
	\begin{subfigure}[t]{1\textwidth}
		\centering
		\includegraphics[height=7.5cm]{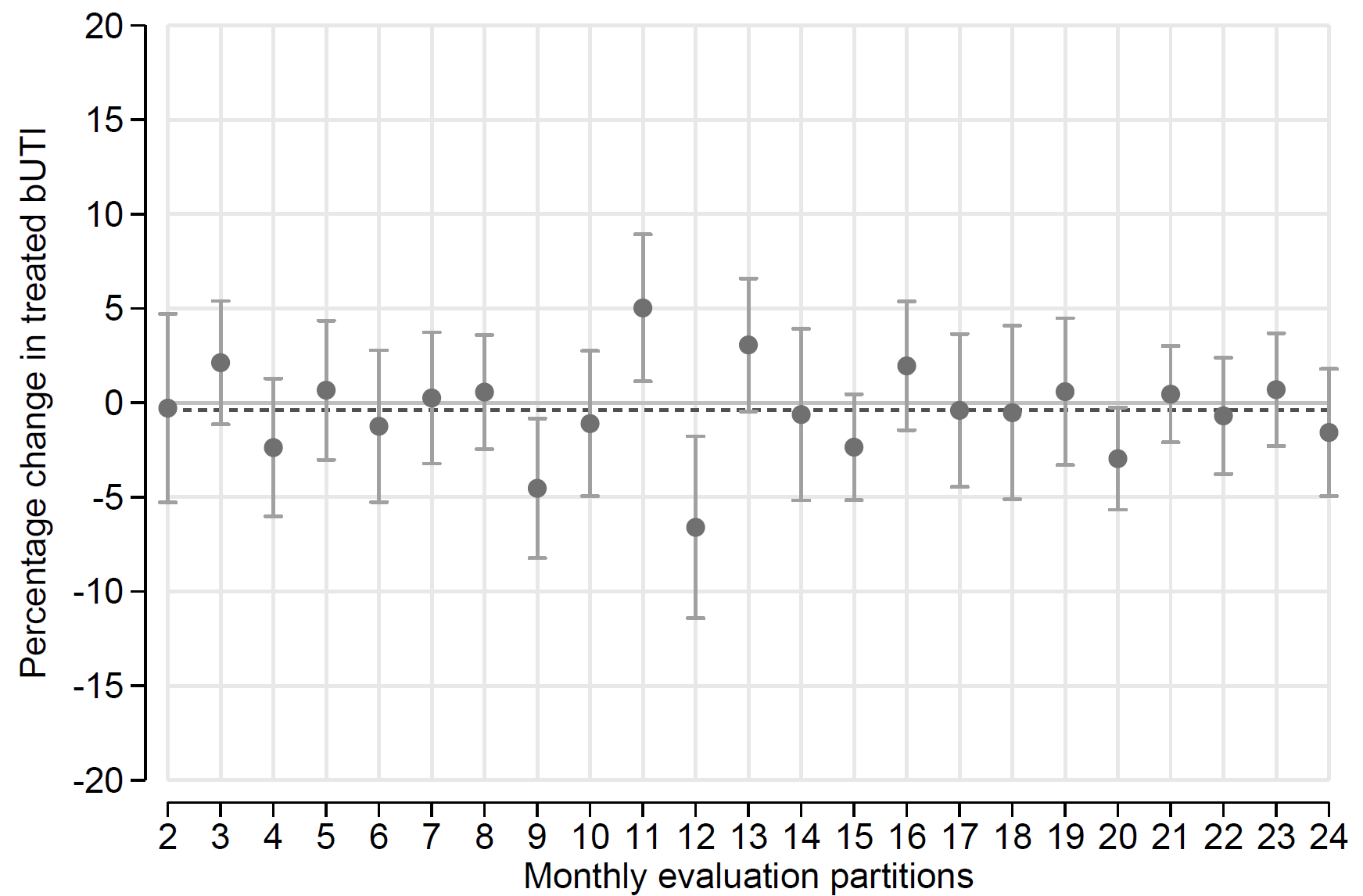}
		\caption{Constraint on treated bacterial UTIs}
	\end{subfigure}
	\caption{\emph{Ex ante} policy outcome for the antibiotic reduction rule by monthly evaluation periods from February 2011 to December 2012. 95\% confidence intervals are computed by bootstrapping the evaluation periods 100 times. The dashed lines are the average outcomes across all evaluation periods combined.}
	\label{fig:ex_ante_ab_outcomes}
\end{figure*}

\begin{figure*}
	\centering
	\begin{subfigure}[t]{1\textwidth}
		\centering
		\includegraphics[height=7.5cm]{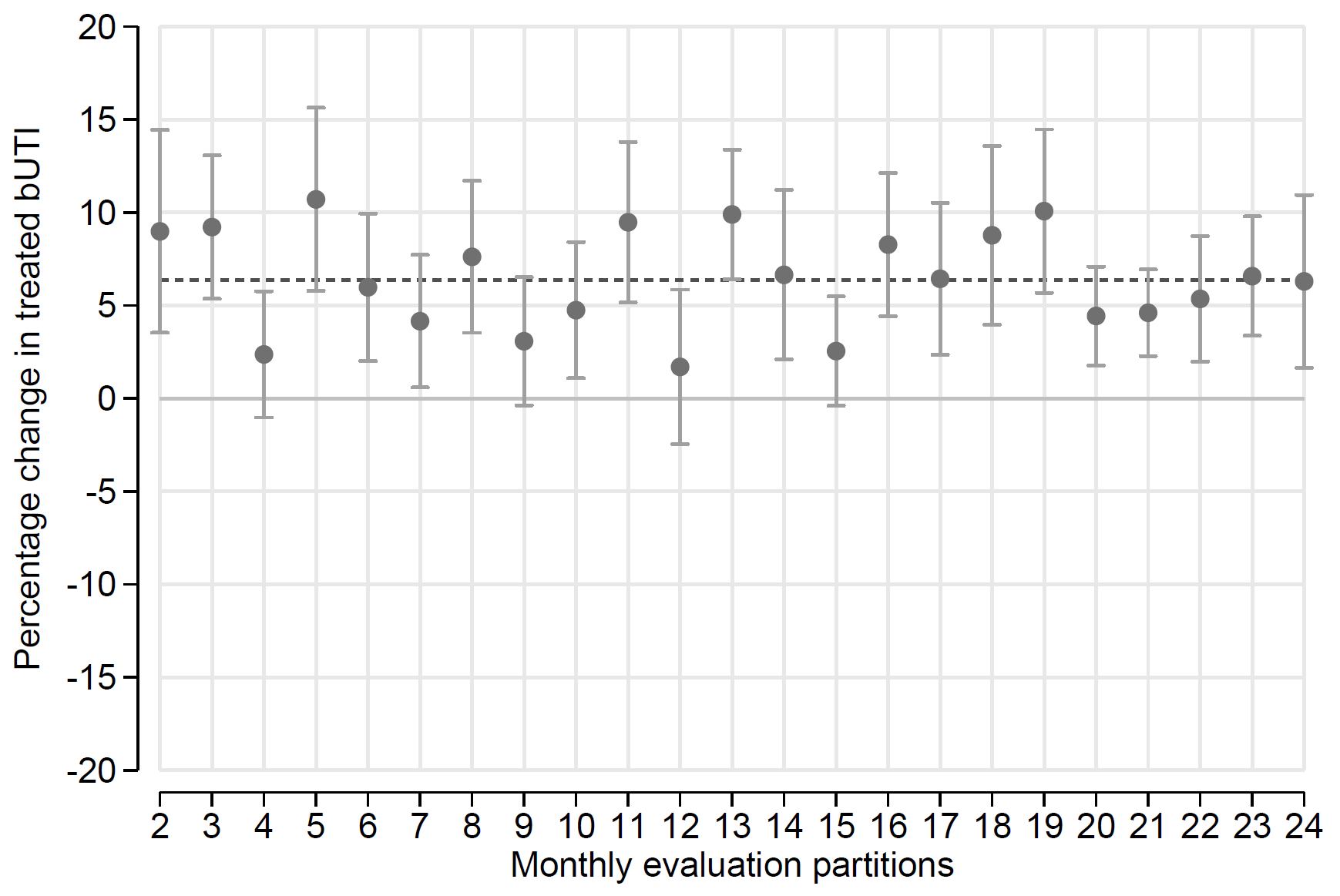}
		\caption{Increase in treated bacterial UTIs}
	\end{subfigure}\\
	\vspace{1cm}
	\begin{subfigure}[t]{1\textwidth}
		\centering
		\includegraphics[height=7.5cm]{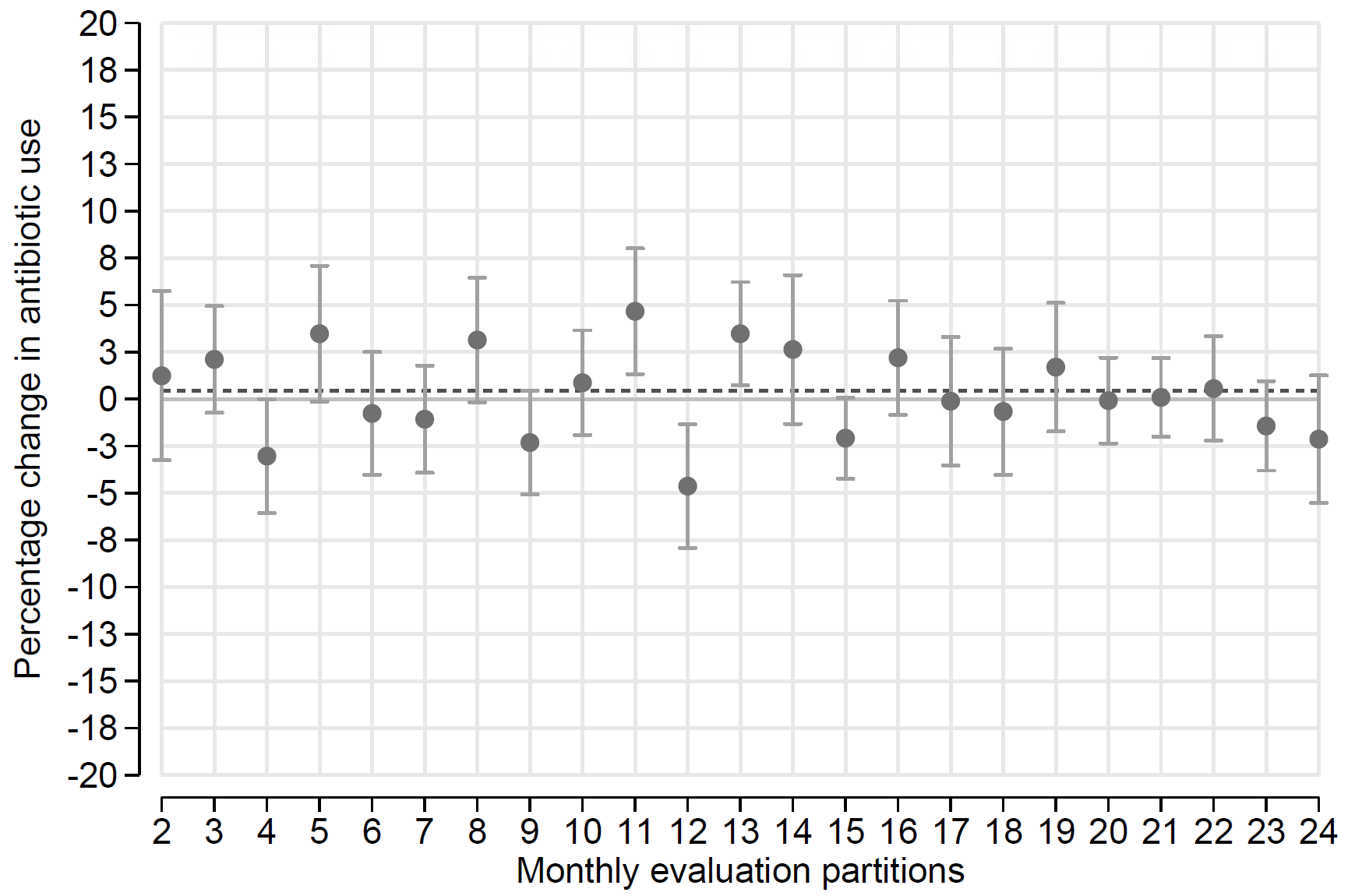}
		\caption{Constraint on antibiotic use}
	\end{subfigure}
	\caption{\emph{Ex ante} policy outcome for the increase in treated bacterial UTI rule by monthly evaluation periods from February 2011 to December 2012. 95\% confidence intervals are computed by bootstrapping the evaluation periods 100 times. The dashed lines are the average outcomes across all evaluation periods combined.}
	\label{fig:ex_ante_bact_outcomes}
\end{figure*}

In our application, we set $\tau$ equal to one month, $\alpha$ to 0.8, and $\lambda$ to one week. The \emph{ex ante} policy outcomes are shown at the monthly level in Figure \ref{fig:ex_ante_ab_outcomes} and Figure \ref{fig:ex_ante_bact_outcomes}, where we continue to number the evaluation periods as in Figure \ref{fig:ex-post-policy}. For the antibiotic reduction objective, shown in Figure \ref{fig:ex_ante_ab_outcomes}, we achieve reductions in antibiotic use significantly different from zero for each month with the exception of November 2011 (evaluation period 11). The changes in the number of treated bacterial UTIs are close to zero and confidence intervals cover the zero in 19 of the 23 implementations. In Figure \ref{fig:ex_ante_bact_outcomes}, the results for the increase in treated bacterial UTI rule are similar. Table \ref{tab:ex-ante-policy} reports aggregate \emph{ex ante} policy outcomes for all evaluation periods. The attained reduction in total antibiotic prescribing of 7.42 percent is less than the \emph{ex post} 9.57 percent but the tight confidence intervals suggest that the improvement is robust.

The same is true for the attained increase in treated bacterial UTI, which is 6.38 percent relative to the \emph{ex post} achievable 7.87 percent. For both policy objectives, the constraints are close to zero and the confidence intervals cover the zero in both cases. These results show that we can achieve both prospective reductions in antibiotic prescribing and an increase in correct treatment of bacterial UTI using machine learning predictions and policies designed based on administrative data.

\begin{table}[h!]
\centering
\small
\begin{threeparttable}
\caption{\emph{Ex ante} robust policy outcomes}
\label{tab:ex-ante-policy}
\begin{tabular}{ @{} l @{\extracolsep{10mm}} c @{\extracolsep{6mm}} c @{} } \\[-1cm]
\toprule
				& \multicolumn{2}{c}{Percentage change in} \\
				\cline{2-3}
Policy objective 	& Antibiotic use 	& Treated bacterial UTIs \\
\midrule
Antibiotic reduction 		& $-7.42 \ \left[-8.07, -6.77\right]$ & $-0.39 \ \left[-1.11,\ 0.34\right]$ \\
Increase treated bacterial UTI 	& $0.43 \ \left[-0.23, 1.09\right]$ & $6.38 \ \left[5.59,\ 7.18\right]\ $ \\
\bottomrule
\end{tabular}
\vspace{-0.2cm}
\footnotesize 95\% confidence intervals are computed based on 100 bootstrap samples.
\end{threeparttable}
\end{table}

\subsection{Are physicians needed?}
\label{algostepfunction}
\begin{figure*}[h!]
	\centering
	\begin{subfigure}[t]{0.46\textwidth}
		\centering
		\includegraphics[height=7cm]{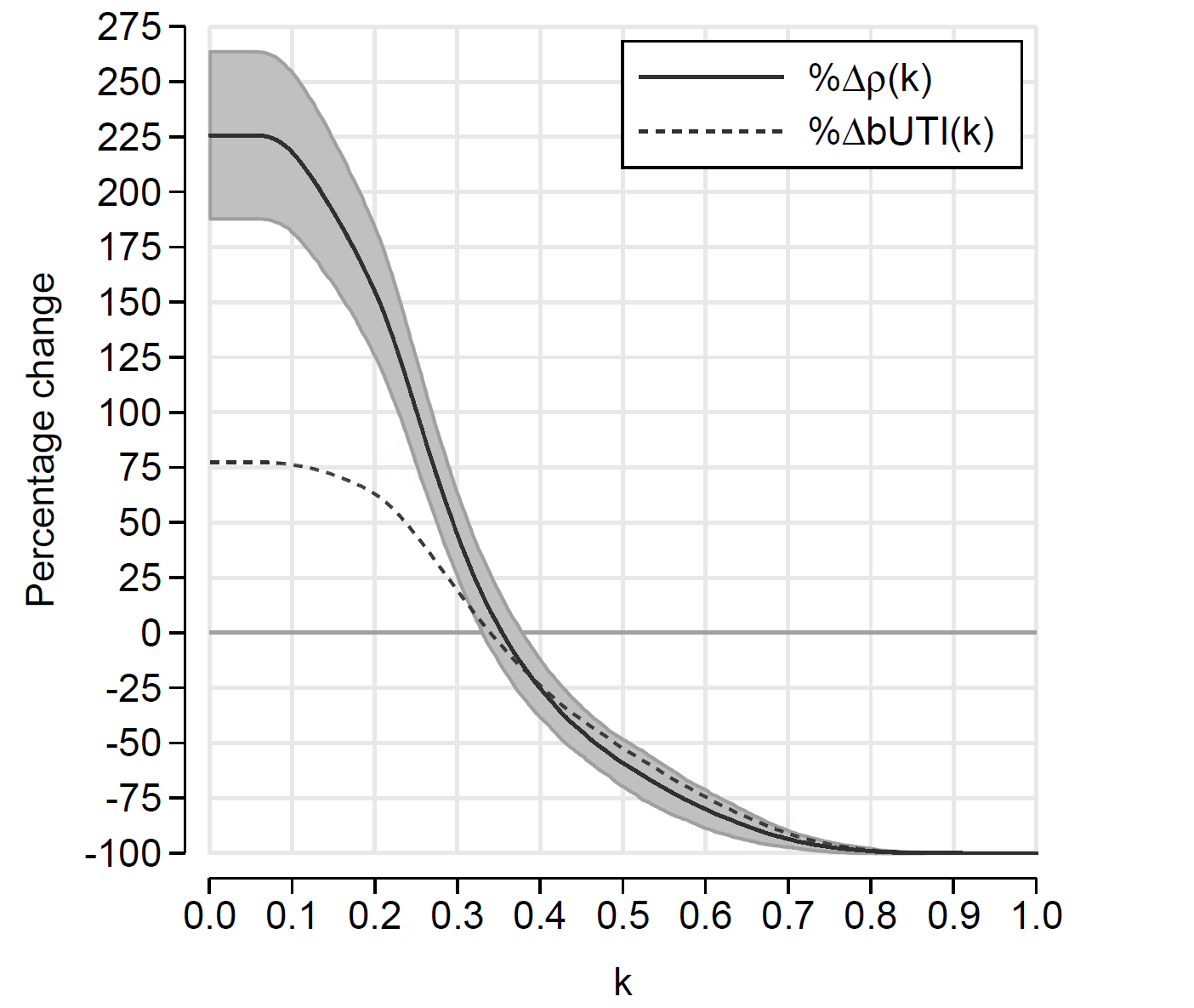}
		\caption{Antibiotic reduction}
	\end{subfigure}
	\hspace{0.5cm}
	\begin{subfigure}[t]{0.46\textwidth}
		\centering
		\includegraphics[height=7cm]{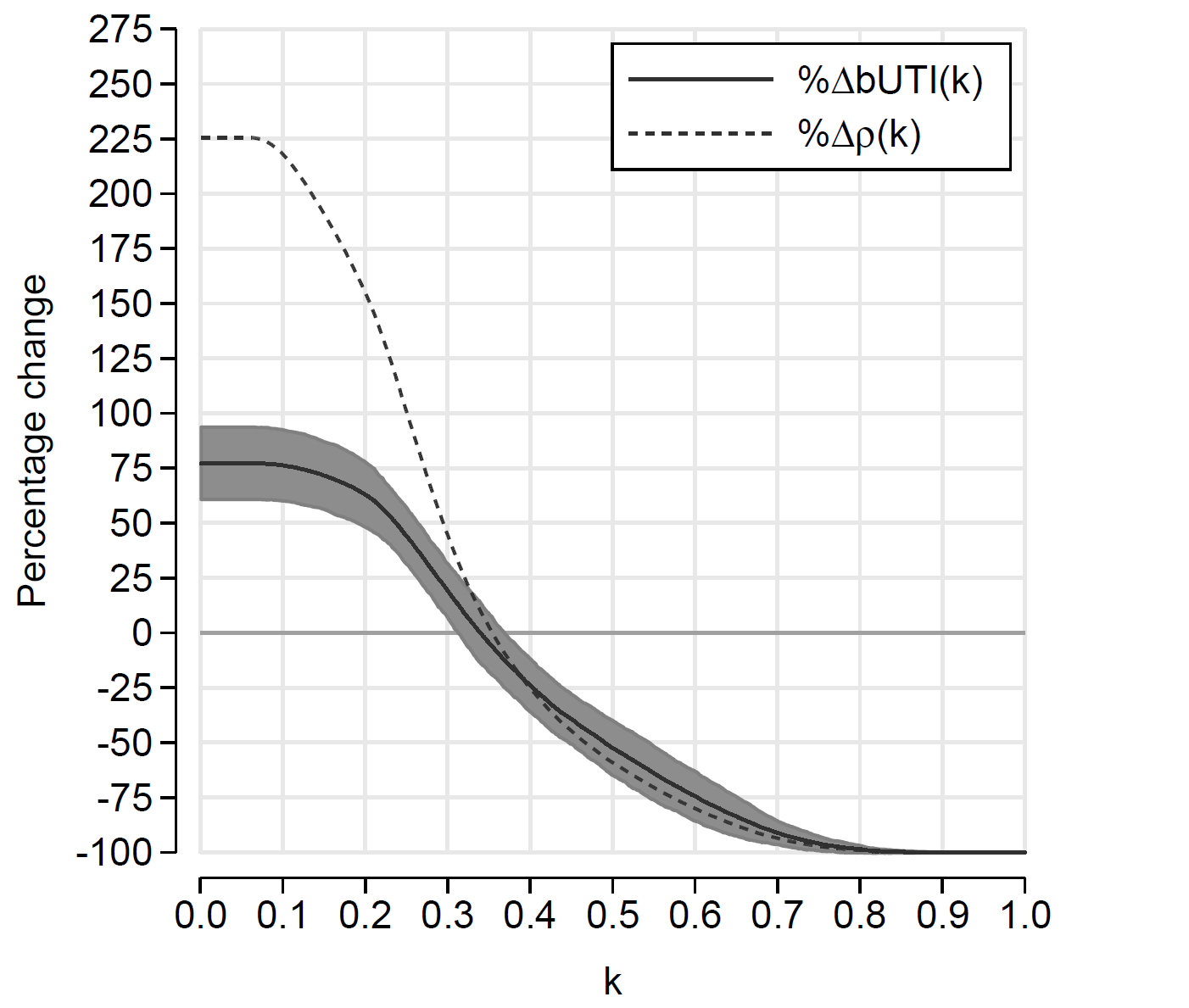}
		\caption{Increase in treated bacterial UTIs}
	\end{subfigure}
	\caption{Prediction-based redistribution policy excluding physician expertise. Outcomes are averaged across evaluation periods and plotted with 95\% confidence intervals.}
	\label{fig:step_function_machine_only}
\end{figure*}

As a special case of the prescription rule in equation (\ref{eqn:general_stewardship_rule}) we can set $k_{L}=k_{H}$. In this case prescription policies are based entirely on algorithm predictions, ignoring physician decision making. The prescription rule becomes a step function where prescriptions are never given below the cut-off $k$ and always given above $k$:
\begin{equation}
	\rho^{M} \left( m(x) ; k \right) \;\; = \;\;
		\begin{cases}\\[-1cm]
			\;\; 0		& \hspace{0.4cm} \text{if } m(x) < k, \\
			\;\; 1		& \hspace{0.4cm} \text{if } k \leq m(x),
		\end{cases}
\label{eqn:machine_only_rule}
\end{equation}
Figure \ref{fig:step_function_machine_only} shows the average outcome from implementations of prescription rule $\rho^{M}(k)$ across all evaluation periods for all possible $k$. We find that the outcomes are inferior to policy rules that include physician expertise. This is indicated by the intersects with zero such that it is impossible to set any $k$ to attain a reduction in antibiotic use and an increase in the number of treated bacterial UTIs.

In particular, the rules in equation (\ref{eqn:machine_only_rule}) fail to deliver improvements regardless of policy maker preferences, $a$ and $b$. For $k=0$, every patient receives a prescription. Then, prescribing increases on average by 226 percent while increasing the number of correctly treated bUTIs by only 77 percent. At the other extreme, $k=1$, no patients are treated at the initial consultation. Therefore, 100 percent of the initially prescribed antibiotics are not given while 100 percent of the patients with actual bacterial UTIs are untreated. For interior $k$, $\rho^{M}$ cannot reduce average antibiotic use, $\Delta \rho(k)$, without also decreasing the number of treated bacterial UTIs. Equivalently, $\rho^{M}$ cannot increase the number of treated bacterial UTIs, $\Delta \text{bUTI}$, without also increasing antibiotic use. Checking individual evaluation periods separately, we find a break even only for evaluation period four on a narrow set of $k$. In all other evaluation periods $\rho^{M}$ fails across the full $k$-range. We conclude that an algorithm decision rule alone cannot ensure improvements. Even with high-dimensional individual-specific data, machine learning predictions need to be combined with physician expertise.

\section{Discussion}
\subsection{Patient heterogeneity in sickness cost}
\label{discuss:sickcost}



We show that machine prediction-based prescription policies can substantially reduce over- and underprescribing. Yet, an alternative explanation for discrepancies between the physicians' prescription choices and the algorithm's decision rule is the potential misspecification of physician preferences. Until now, we ignored one potentially problematic assumption in our model: sickness cost, $a_{S}$, is constant across all patients in the policy maker's utility specification. Some patients might be high cost, not to be confused with high risk, in the sense that the health consequence of having a bacterial UTI would be much more severe for such patients compared to others. One important example are pregnant women. For example, Schieve et al. (1994) show that UTIs during pregnancy are positively associated with adverse outcomes such as low birthweight, prematurity, maternal anemia, and others. Foxman (2002) in a review reports that a UTI during pregnancy leads to elevated risk of kidney infection, fetal mortality, and premature delivery. Schwandt (2018) finds that in-utero exposure to influenza infections leads to premature delivery and low birthweight in the short-run and a nine percent reduction in earnings and a 35 percent increase in welfare dependence in the long-run. 

If physicians immediately prescribe to pregnant women, not because they fail to properly evaluate risk, but rather because they assign a high sickness cost, it is then indeed optimal to prescribe even at low risk of bacterial infection. We can identify patient pregnancy in our data by claims for periodic examinations during pregnancy required by law. Pregnant women account for 28 percent of patients in the hold-out sample.\footnote{Pregnant women have higher risk of UTI and guidelines recommend that physicians always acquire urine culture from pregnant women showing UTI symptoms. Combined with a higher sickness cost, it is not surprising that they are overrepresented in our data (Foxman, 2002).} To examine if our policy results are driven by redistributing away from low risk but high cost pregnant women, we implement a modification to the prescription rule $\rho^{S}$ in that physicians' prescriptions to pregnant woman cannot be changed.
\begin{table}[h]
\centering
\small
\begin{threeparttable}
\caption{\emph{Ex ante} policy outcomes excluding pregnant women}
\label{tab:ex-ante-policy-nopreg}
\begin{tabular}{ @{} l @{\extracolsep{10mm}} c @{\extracolsep{6mm}} c @{} }\\[-1cm]
\toprule
				& \multicolumn{2}{c}{Percentage change in} \\
				\cline{2-3}
Policy objective 	& Antibiotic use 	& Treated bacterial UTIs \\
\midrule
Antibiotic reduction 		& $-6.81\ \left[-7.73,\ -5.88\right]$ & $0.61 \ \left[-1.79,\ 0.57 \right]$ \\
Increase treated bacterial UTI 	& $0.36\ \left[-0.53,\ 1.25\right]$ & $5.56 \ \left[4.59,\ 6.53\right]\ \ $ \\
\bottomrule
\end{tabular}
\vspace{-0.2cm}
\footnotesize 95\% confidence intervals are computed based on 100 bootstrap samples.
\end{threeparttable}
\end{table}

Table \ref{tab:ex-ante-policy-nopreg} reports the results of redistributing prescriptions according to our policy rules but excluding all pregnant women. The main results remain qualitatively unaffected as the confidence intervals largely overlap for the full data and the data excluding pregnant women. This finding indicates that physicians do not tend to base their prescription decisions on higher sickness cost for pregnant women but their evaluation of risk. Therefore, we conclude that our are not diminished by redistributing prescriptions away from high sickness cost patients.

\subsection{How well does the policy assign prescriptions?}
\label{discuss:assigntreatment}

We provide some indication that our policy is unlikely to prevent antibiotic treatment of high sickness cost patients. Given the policy objective function relies on redistributing antibiotics from low to high predicted risk patients, we may also worry that antibiotics are given unnecessarily, even if predicted bacterial risk is high. For example, asymptomatic infections are often left untreated. In addition, our policy considers prescriptions given before the test results become available. Once the laboratory test results arrive several days later, treatment choices can be made under full information. Therefore, physician may have a very accurate prior for high risk patients but decide to postpone treatment until the test results are available. This may be feasible if symptoms are bearable or if prescription-free pain medication is used in the meantime.

To investigate whether the policy gives antibiotics to high predicted risk patients who would not be treated even under full information about the presence of bacteria in their sample, it is informative to consider physicians' prescription choices with full information, after the arrival of test outcome. We find that, out of the predicted high risk patients to whom physicians did not prescribe an antibiotic before knowing the test result, 70\% received an antibiotic within 10 days of the initial consultation. Considering the spontaneous recovery rate of 24\% reported in Ferry et al. (2004), this finding gives us confidence that patients are typically treated with antibiotics when bacterial presence in a urine sample is known with certainty. The redistribution policy based on machine predicted risk hence truly increases treated bacterial infections early on.

One further reason could lead physicians to postpone treatment: lacking information about antibiotic resistance. Even if the physician knows with high accuracy that a patient suffers from a bacterial infection, the bacterial species and its resistance profile towards available antibiotics is likely known with less certainty. To avoid prescribing an ineffective treatment, the physician may choose to wait for the test results, even if predict bacterial risk is high. This would imply that the information about high predicted bacterial risk is not useful if no information about bacterial species and resistance can be provided. Predicting these is a more complex matter that we do not consider here. Yet, to gain a better understanding of the importance of this potential reason for postponing treatment, we consider resistance profiles of bacterial species found in high predicted risk patients, conditional on physician prescription decisions. If physicians know with high accuracy whether an infection is bacterial and suspect resistance to be high, we should observe that resistance is higher for bacteria found in patients for whom physicians did not give a prescription at initial consultations than for bacteria found in patients receiving a prescription instantly. 

We observe some differences in the detected bacteria for treated and untreated patients with high predicted bacterial risk. When physicians decide to treat instantly, E.coli account for 74\%, K.pneumoniae for 6\%, Enterococcus for 4\%, and E.faecalis for roughly 3\% of bacteria found. These are the main four bacteria responsible for UTI in this group. When physicians decide to wait, E.coli account for 59\%, K. pneumoniae for 9\%, E. faecalis for 8\%, and Enterococcus for 5\%. The significant difference in the fraction of detected E.coli bacteria can be explained by in-clinic diagnostics done by some physicians, which can identify E.coli bacteria relatively well. Therefore, E.coli bacteria can be treated correctly more frequently at an initial consultation. We also find small but relevant differences in resistances against the bulk of antibiotics prescribed for UTI. When physicians decide to treat instantly and bacteria are found, these are resistant against mecillinam (J01CA11) in 54.86\%, sulfamethizole (J01EB02) in 47.81\%, nitrofurantoin (J01XE01) in 26.64\%, trimethoprim (J01EA01) in 24.87\%, and ciprofloxacin (J01MA02) in 10.91\% of cases. When physicians decide to wait and bacteria are found, these are resistant against mecillinam in 60.47\%, sulfamethizole in 57.96\%, nitrofurantoin in 24.85\%, trimethoprim in 31.50\%, and ciprofloxacin in 13.80\% of cases. Hence, physicians likely have informative priors about levels of antibiotic resistance and they seem to consider them when deciding to treat instantly or to wait in anticipation of complete test results. Given these differences, there seems to be value to address prediction of specific bacteria and resistances in further research. Yet, the differences are moderate so that the decision to postpone seems to be largely driven by uncertainty about bacterial infection causes more generally.

\section{Conclusion}

In this paper, we have shown that policies based on machine learning predictions using administrative data can improve antibiotic prescribing in economically significant ways while maintaining overall treatment quality. Antibiotic prescribing under uncertainty about the cause of infection is a high impact policy problem given the empirical relevance of increasing antibiotic resistance due to wasteful antibiotic prescribing. While we show that implementing policies leads to improvements over physicians' prescription decisions, machine learning predictions alone cannot achieve improvements. Physician expertise, the observation and assessment of factors unobservable to the machine learning algorithm, remains crucial in prescription choices. Still, for a large interval of predicted risks, using machine learning leads to a reduction in over- and underprescribing.

One promising avenue for further research is the combination of machine learning predictions with further clinical information such as recorded symptoms and results from in-clinic diagnostics for bacterial infection causes. Because we omit an important dimension of antibiotic prescribing, the choice of molecule, it remains an open question for future research to what extent machine prediction of bacterial species and molecule-specific resistances are able to further inform prescription choices. An important area in which further research is needed is the analysis of experts' behavioral reactions to the prediction-based policies proposed here as physicians' incentives to treat and test are likely to change. Finally, our machine prediction results could be used to assist physicians in their decision-making, for example by providing physicians with a machine predicted risk at every prescription occasion. A full assessment of the equilibrium effects of such an assistance will require interventions in the field combined with machine prediction, a promising avenue for future research to further improve antibiotic prescribing.

One limitation is that we consider only first consultation prescription occasions in which a laboratory test was ordered. Our analysis still holds empirical relevance because UTI typically must be treated quickly and laboratory testing takes considerable time. Ribers and Ullrich (2018) consider a more general setting capturing observed prescription choice occasions throughout complete treatment spells in a structural dynamic model that endogenizes both the prescription and diagnostic test decision. Extending the machine learning approach considered here to a similarly general setting is beyond the scope of this paper. Randomizing additional laboratory testing could allow for machine learning predictions to be evaluated against physician decisions on a generalizable sample of the population. 

Our analysis shows the potential of machine learning methods for policies in specific healthcare treatment choice situations. The quality of prediction algorithms and data availability are improving at a rapid pace. Yet, the challenge of evaluating machine learning predictions against human expert decisions will remain in the presence of unobservables, decision-makers' heterogenous technologies, and moving objective functions. These challenges likely need to be carefully solved taking into account domain-specific context in each implementation setting.


\newpage
\appendix

\newpage
\section{Machine learning performance: feature importance}
\label{app:RF}

\begin{figure*}[h!]
	\centering
	\includegraphics[height=8.5cm]{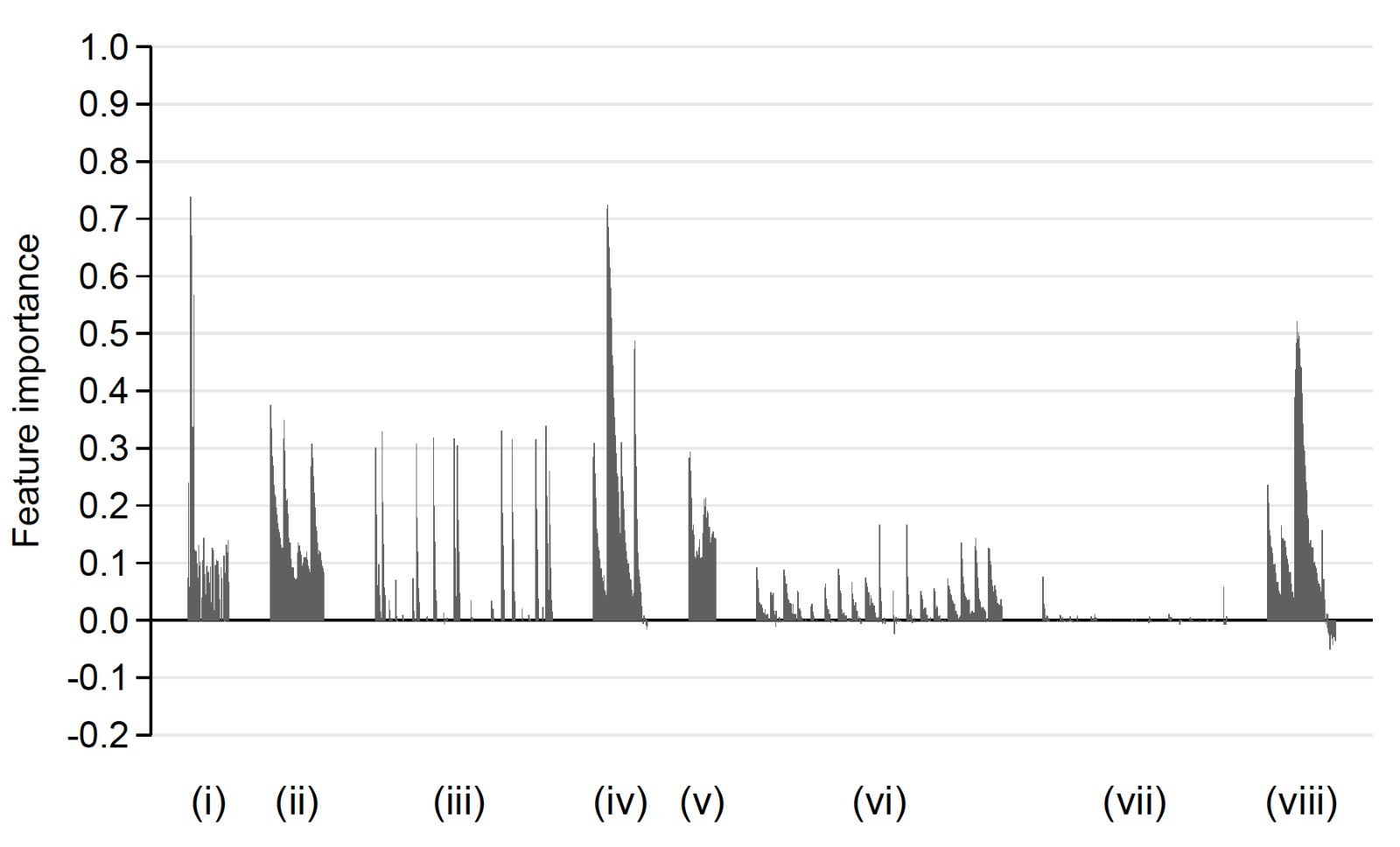}
    	\caption{Feature importance averaged over the random forests used on each of the 24 monthly folds from January 2011 to December 2012. Variables are listed by groups containing (i) patient characteristics and test timing; (ii) patient past prescriptions; (iii) patient past resistance test results; (iv) patient past hospitalizations; (v) patient past general practice insurance claims; (vi) household members' past prescriptions; (vii) household members' past resistance test results; and (viii) household members' past hospitalizations.}
	 \label{fig:feature-importance}
\end{figure*}

\newpage
\section{Physician heterogeneity}
\label{app:gp_meandev}

\begin{table}[h]
\centering
\small
\begin{threeparttable}
\caption{Use of unobservables and clinic characteristics}
\label{tab:gp_meandev}
\begin{tabular}{ @{} l @{\extracolsep{10mm}} r @{\extracolsep{3mm}} l @{} }
\toprule
Outcome: mean deviation of bacterial rates	 & \multicolumn{2}{c}{$\beta$, linear regression}  \\
between treatment decisions				 & 	 \\
\midrule
Number of clinic's unique patients per physician 	& 0.002& [-0.0001, 0.005] \\
Number of laboratory results per unique patient  	& 0.079& [-0.044, 0.201]  \\
Mean number of physicians  					& 0.003& [-0.003, 0.008]  \\
Mean age of physicians  						& -0.001& [-0.002, -0.0005]  \\
Share of female physicians  					& 0.0002& [-0.016, 0.017]  \\
\bottomrule
\end{tabular}
\begin{tablenotes}
\item \footnotesize Notes: 95\% confidence intervals (heteroskedasticity-robust) reported in brackets.
\end{tablenotes}
\end{threeparttable}
\end{table}

\clearpage
\section{Constraints under \emph{ex post} policy evaluation}
\label{app:ex-post-violation}

\begin{figure*}[h!]
	\centering
	\begin{subfigure}[t]{1\textwidth}
		\centering
		\includegraphics[height=7.5cm]{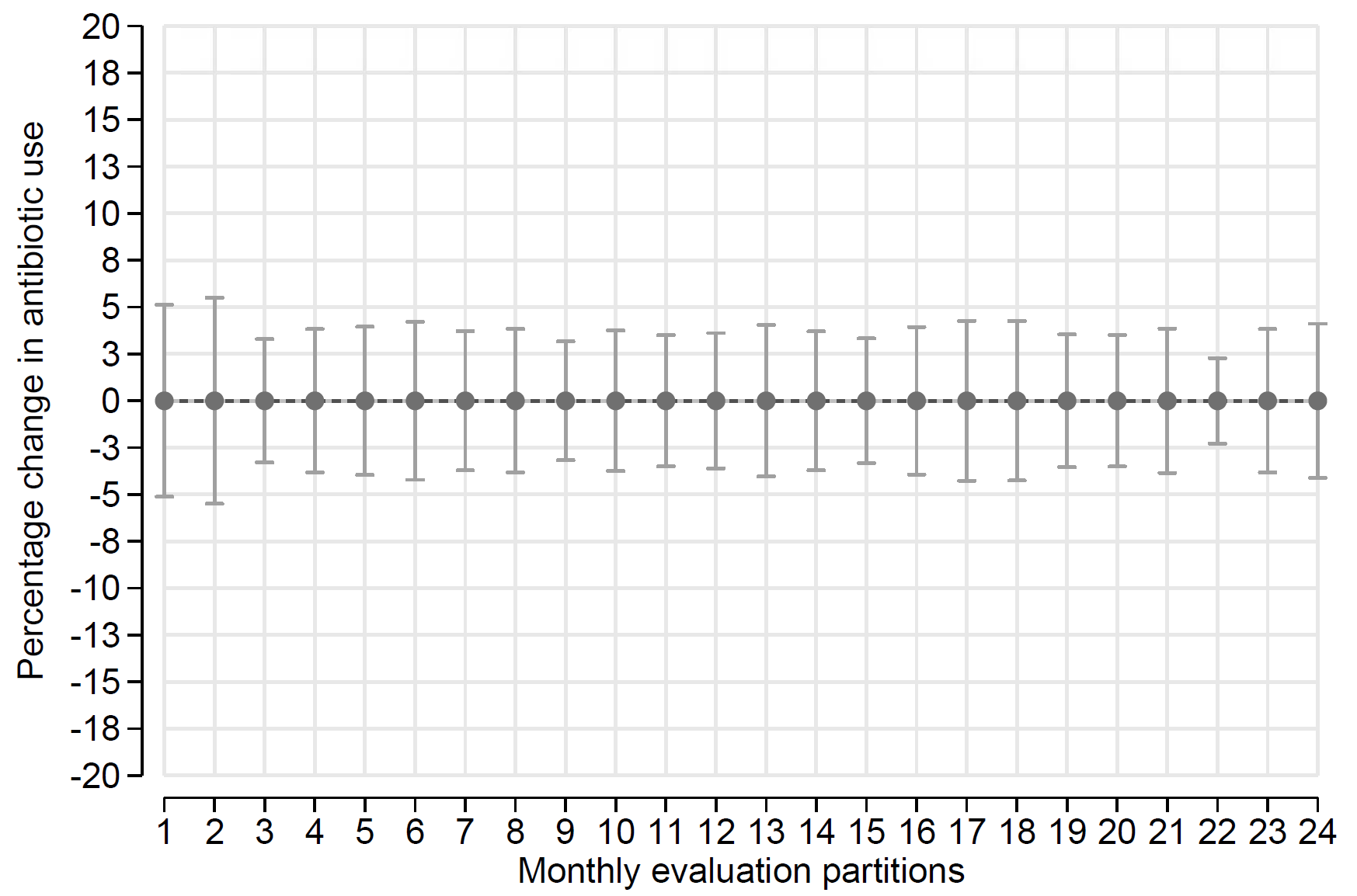}
		\caption{Antibiotic constraint under increase treated bUTI policy}
	\end{subfigure}\\
	\vspace{1cm}
	\begin{subfigure}[t]{1\textwidth}
		\centering
		\includegraphics[height=7.5cm]{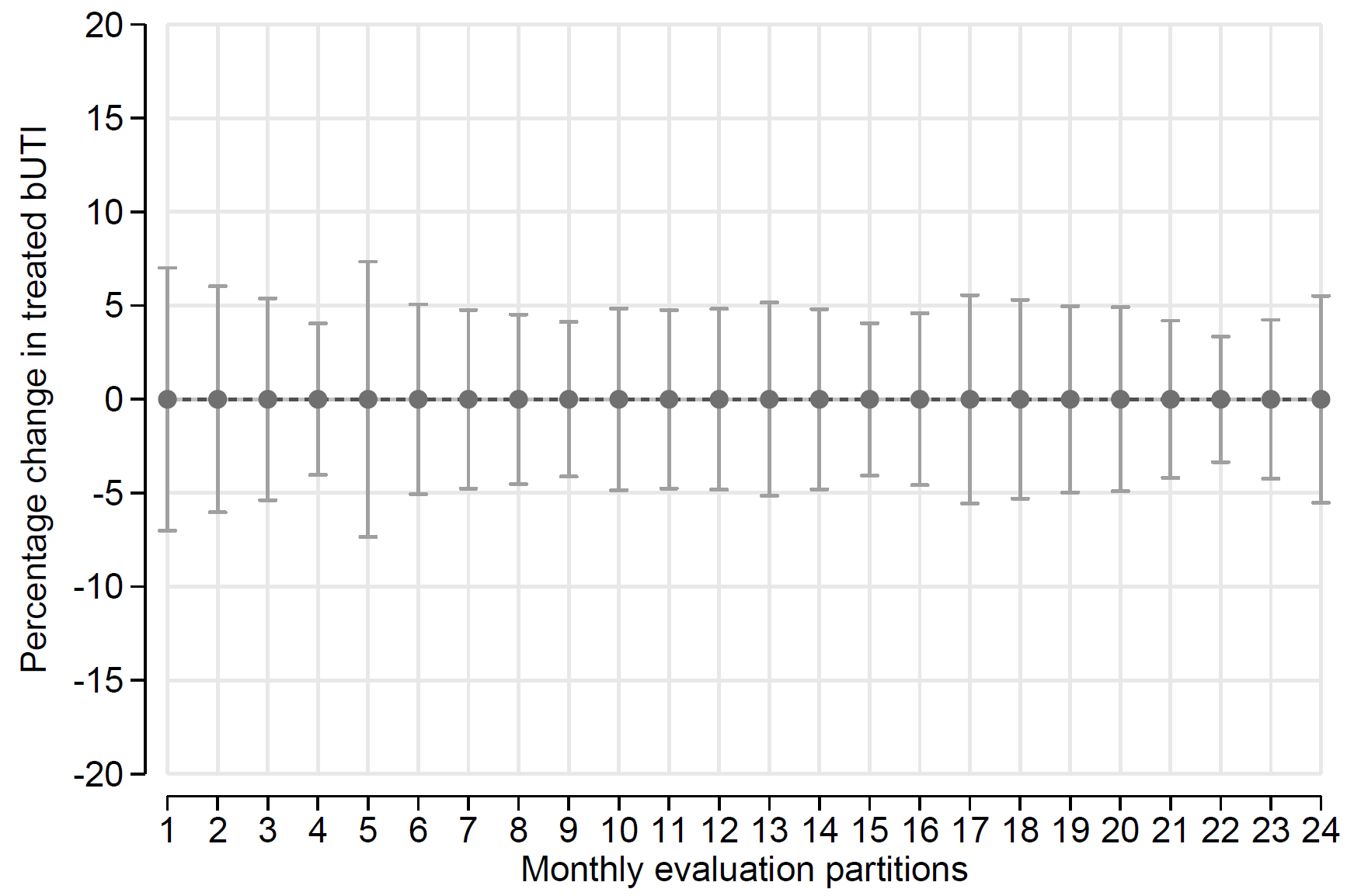}
		\caption{Treated bacterial UTI constraint under antibiotic reduction policy}
	\end{subfigure}
	\caption{Constrains for the \emph{ex post} policy rule by monthly evaluation periods from January 2011 to December 2012. 95\% confidence intervals are computed by bootstrapping the evaluation periods 100 times. The dashed line is the outcome for all evaluation periods combined.}
	\label{fig:ex-post-policy_constraint}
\end{figure*}

\end{document}